\documentclass[12pt]{article}
\usepackage{epsfig, amssymb}
\usepackage{graphicx,psfrag} 
\newcommand{\be}{\begin{equation}}
\newcommand{\ee}{\end{equation}}
\newcommand{\bea}{\begin{eqnarray}}
\newcommand{\eea}{\end{eqnarray}}
\newcommand{\bA}{\begin{array}}
\newcommand{\eA}{\end{array}}
\newcommand{\bc}{\begin{center}}
\newcommand{\ec}{\end{center}}
\newcommand{\al}{\alpha}

\newcommand{\ra}{\rightarrow}
\newcommand{\del}{\partial}

\newcommand{\ie}{{\it i.e.}}
\newcommand{\eg}{{\it e.g.}}

\newcommand{\C}{\mathbb{C}}

\newcommand{\Rea}{\mathop{\rm Re}}

\newcommand{\Nt}{${\cal N}{=}2$}

\def\BC{{\mathbb C}}
\def\BP{{\mathbb P}}
\def\BR{{\mathbb R}}
\def\BZ{{\mathbb Z}}
\def\BQ{{\mathbb Q}}

\def\BN{\mbox{\boldmath$N$}}

\begin{document}

\begin{titlepage}
\vspace{30mm}

\bc

\hfill  {\tt arXiv:0912.3374 [hep-th]} 
\\         [22mm]

{\huge On nonsupersymmetric $\BC^4/\BZ_N$, tachyons, \\ [2mm]
terminal singularities and flips}
\vspace{16mm}

{\large K.~Narayan} \\
\vspace{3mm}
{\small \it Chennai Mathematical Institute, \\}
{\small \it SIPCOT IT Park, Padur PO, Siruseri 603103, India.\\}

\ec
\medskip
\vspace{30mm}

\begin{abstract}
We investigate nonsupersymmetric $\BC^4/\BZ_N$ orbifold singularities
using their description in terms of the string worldsheet conformal
field theory and its close relation with the toric geometry description 
of these singularities and their possible resolutions. Analytic and
numerical study strongly suggest the absence of nonsupersymmetric
Type II terminal singularities (\ie\ with no marginal or relevant
blowup modes) so that there are always moduli or closed string
tachyons that give rise to resolutions of these singularities,
although supersymmetric and Type 0 terminal singularities do exist. 
Using gauged linear sigma models, we analyze the phase structure of 
these singularities, which often involves 4-dimensional flip 
transitions, occurring between resolution endpoints of distinct 
topology. We then discuss 4-dim analogs of unstable conifold-like 
singularities that exhibit flips, in particular their Type II GSO 
projection and the phase structure. We also briefly discuss aspects 
of M2-branes stacked at such singularities and nonsupersymmetric 
$AdS_4\times S^7/\BZ_N$ backgrounds.
\end{abstract}

\end{titlepage}

\newpage 
{\footnotesize
\begin{tableofcontents}
\end{tableofcontents}
}

\vspace{2mm}

\section{Introduction and summary}

The interplay between string theory and geometry is a fascinating
subject, in many cases beautifully elucidated by gauged linear sigma
models \cite{wittenphases}, exhibiting rich structures such as
topology change and hints of a quantum completion of classical
geometry (see \eg\ \cite{greeneCY,wittenIAS} for reviews). This also
turns out to be remarkably useful in cases where spacetime
supersymmetry is broken, as in the context of unstable geometries and
their resolution via closed string tachyon condensation.  It was
argued in \cite{aps} using the geometry seen by D-brane probes that
unstable $\BC/\BZ_N$ singularities get smoothed out or resolved by the
condensation of closed string tachyons localized at the singular
tips. The physical picture here was also shown to hold for 2-dim
$\BC^2/\BZ_N$ singularities in \cite{aps,vafa,hkmm}, (see
\cite{emilrev,minwalla0405} for reviews) as well as 3-dim
$\BC^3/\BZ_N$ and conifold-like singularities
\cite{drmknmrp,drmkn,knconiflips}, where string worldsheet
renormalization group flows were used (see also \cite{sarkar0407}). 
This is due essentially to the existence of $(2,2)$ worldsheet
supersymmetry which protects various observables along renormalization
group flows. Although these first order renormalization group flow
equations are not quite the same as time evolution in spacetime (being
described by second order equations), various qualitative features,
such as the directions of evolution of the geometry and the structure
of fixed points, are in fact robust.  An elegant description of the
physics here is obtained from renormalization group flows in closely
related gauged linear sigma models (GLSMs), which in turn dovetail
nicely with the description of singularity resolution in algebraic
geometry in the context of nonsupersymmetric 2- and 3-dim
singularities.

In this paper, we study nonsupersymmetric $\BC^4/\BZ_N$ and
conifold-like singularities (which are both toric) using the
techniques developed for the 3-dim case. This is of interest both from
the point of view of understanding string and M-theory backgrounds
constructed using such complex 4-dim spaces with singularities, as
well as understanding the geometric structure of closed string tachyon
condensation in 4-dim spaces, which is somewhat richer than the lower
dimensional cases.  In recent times, we have seen the
emergence of an understanding of the dual field theories to M-theory
$AdS_4\times S^7/\BZ_N$ backgrounds obtained from the near horizon
limits of M2-branes stacked at supersymmetric $\BC^4/\BZ_N$ orbifold
singularities \cite{abjm}: see also closely related work \cite{bgl}. 
Various generalizations of this, in particular with M2-branes stacked 
at other 4-dim singularities, have been studied in \eg\
\cite{klebanov,terashima,jafferis,imamura,uedayama,lee,ms1,ms2,
kleb2,hanany1,hanany2,hesparks} (see also \eg\
\cite{morrisonplesserHorizons,zaffaroni} for some early work on 4-dim
singularities). This gives additional perspective to the
nonsupersymmetric case we deal with here.

The backgrounds we consider here are of the form M-theory on
$\BR^{2,1}\times\BC^4/\BZ_N$, or Type IIA on
$\BR^{1,1}\times\BC^4/\BZ_N$ obtained by compactifying one of the
directions in $\BR^{2,1}$ on a circle. The basic fact we exploit
towards obtaining insight into the physical structure of these
singularities is that the Type II closed string blowup modes of the
orbifold singularity (equivalently the complexified Kahler parameters
of its various collapsed cycles) map to metric and 3-form modes when
lifted to M theory. To elaborate further, as in the investigation
\cite{gukovvafawitten} of M-theory on Calabi-Yau 4-folds (see also
\cite{gukovgateswitten}), the classical Kahler parameters arising from
variations of the metric combine with scalars dual (in $\BR^{2,1}$) to
the $U(1)$ gauge fields that arise from compactification of the 3-form
on $\BC^4/\BZ_N$, to give complex scalars representing tachyons or
moduli in M-theory on the uncompactified $\BR^{2,1}$. Compactifying
one of the $\BR^{2,1}$ directions on a circle gives Type IIA string
theory, with the tachyons or moduli arising as complex scalars from
metric variations and the $B$-field compactified on
$\BC^4/\BZ_N$. These complex scalars govern the geometric blowup modes
of the $\BC^4/\BZ_N$ singularity, in string or M-theory: in string
theory, these are just NS-NS sector modes, with the RR sector playing
no essential role in the resolution of the singularity by tachyon
condensation (as in the lower dimensional cases). The resolution
structure of these singularities, described by the toric geometry of
the singularity, is beautifully captured by gauged linear sigma
models.  A precise correspondence between these geometric blowup modes
and twisted sector string states in the string worldsheet orbifold
conformal field theory was established for nonsupersymmetric 
$\BC^2/\BZ_N$ \cite{hkmm} and $\BC^3/\BZ_N$ \cite{drmknmrp}: this is 
a generalization of the well-known correspondence between \eg\ the 
$N-1$ blowup modes of a supersymmetric $\BC^2/\BZ_N$ (\ie\ ALE) 
singularity and twisted sector string states. We describe and use a 
similar correspondence here to understand the phase structure of
$\BC^4/\BZ_N$ singularities using in part their orbifold conformal
field theory description.  Alternatively, assuming the correspondence
between twisted sector string states and geometric blowup modes is
faithful and complete, our use of the orbifold conformal field theory
as a substitute for the toric geometry and resolution structure of the
singularity can be used to gain insights into M-theory compactified on
such singularities. It would then seem that a similar correspondence
might govern the resolution structure of such singularities as
described by M-theory, and a more direct M-theory analysis of this
would be interesting.

Various aspects of the structure of nonsupersymmetric $\BC^4/\BZ_N$
orbifold singularities are similar to those of $\BC^3/\BZ_N$, with
some notable new features too (some early work in the mathematics
literature on 4-dim quotient singularities appears in \eg\
\cite{morrisonstevens}). These singularities, generically incomplete
intersections, do not admit any complex structure deformations as for
$\BC^3/\BZ_N$ \cite{schless}. They contain localized closed string
tachyons or moduli in their twisted sector spectrum governing the
blowup modes of the singularity (which are all Kahler): these can be
classified into several rings that are chiral and antichiral with
respect to the worldsheet supersymmetry, and respect different complex
structure with respect to the spacetime coordinates. It is possible to
find, for various families of orbifolds, an appropriate Type II GSO
projection that ensures that the only tachyons in the system are these
localized ones. This GSO projection typically preserves some tachyons
in each ring, projecting the others out.
As in $\BC^3/\BZ_N$ orbifolds, condensation due to all tachyons in a
given chiral or antichiral ring does not completely resolve a
$\BC^4/\BZ_N$ singularity, unlike the lower dimensional cases: there
exist ``geometric terminal'' singularities, with no Kahler blowup modes 
(these are what are usually referred to as terminal singularities in 
the mathematics literature). A
new feature in $\BC^4/\BZ_N$ is the existence of supersymmetric
terminal singularities, which do not admit any blowups, Kahler or
nonKahler, at the level of the string worldsheet (or equivalently the
toric geometry).  It was proven in \cite{drmknmrp} that
nonsupersymmetric Type II $\BC^3/\BZ_N$ orbifolds always contain a
tachyon or modulus in some ring, \ie\ they are never (all-ring)
terminal, so that such a singularity will always be resolved by some
Kahler or nonKahler blowup due to tachyon condensation. In this
aspect, the combinatorics of $\BC^4/\BZ_N$ orbifolds appears much more
intricate and an analytical understanding of whether Type II
nonsupersymmetric terminal singularities exist appears
difficult. However using some analytic techniques and some numerical
investigation via a Maple program, it is possible to gain some insight
into the structure of these singularities. Our investigation strongly
suggests that in fact nonsupersymmetric terminal singularities do not
exist, as in $\BC^3/\BZ_N$, \ie\ Type II $\BC^4/\BZ_N$ orbifolds
always contain a tachyon or modulus in their twisted sector spectrum.
This implies that the final endpoints of closed string tachyon
condensation in unstable 4-dim Type II singularities are either smooth
(\ie\ completely resolved) or supersymmetric singularities (which can
be terminal, or of lower dimension). It is not clear if there is an
obvious physical reason for the non-existence of nonsupersymmetric
terminal singularities, so the result, if true, is striking.

Tachyons localized at the singularity signal instabilities of the
system which decays via their condensation to more stable endpoints,
which generically are also unstable. This cascade process continues
and eventually stops when the system has no further instabilities. In
the initial stages of the condensation (in the vicinity of the
singularity), gravitational backreaction is negligible so that
analysing just this process of condensation of the localized tachyonic
modes ignoring other string modes is a good approximation. Our 
analysis of the decay structure of this system using GLSMs ties in
closely with the symplectic quotient construction of the resolution of
these singularities (we will mostly not describe the equivalent
holomorphic quotient construction here). Since
multiple decay channels stemming from multiple tachyons exist (there
is no canonical resolution as in $\BC^3/\BZ_N$), the most likely decay
channel corresponds to condensation of the most dominant tachyon with
the most negative mass-squared in spacetime: on the worldsheet, this
is the most relevant twisted sector operator, belonging in some ring.
Geometrically, condensation of such a tachyon induces a
partial resolution of the $\BC^4/\BZ_N$ singularity, with a bubble
(typically a weighted projective space $w\BC\BP^3$ here) expanding
outwards in (RG) time. Typically there are residual singularities on
this expanding locus which could be geometric terminal, \ie\ terminal
with respect to the complex structure of the ring containing the
condensing tachyon. In this case, a tachyon (or modulus) in some 
other ring will induce a blowup further resolving the system,
consistent with the non-existence of Type II terminal singularities. 
Systems with multiple tachyons generically exhibit flip transitions, 
tachyonic analogs of flops: in $\BC^3/\BZ_N$ orbifolds, this is a 
blowdown of an unstable 2-cycle accompanied by a blowup of another, 
more stable, 2-cycle of distinct topology, occurring when a more 
dominant tachyon condenses during the condensation of some tachyon. 
In $\BC^4/\BZ_N$, flips arise from the blowdowns and blowups of 
cycles of different dimensionality, involving weighted $\BC\BP^2$s 
and $\BC\BP^1$s. Thus in a sense, the topology change here is stronger.

We also investigate 4-dim conifold-like singularities here, 
generalizing \cite{knconiflips}. These are described by 
a $U(1)$ action with charges 
$Q=(\bA{ccccc} n_1 & n_2 & n_3 & -n_4 & -n_5 \eA)$, with $n_i>0$. They 
do not have a manifest conformal field theory interpretation but their 
phase structure can be analysed using GLSMs. Based on the known Type II 
GSO projections for orbifolds and the phase structure which typically 
contains residual orbifold singularities, we guess a Type II GSO 
projection,\ $\sum_iQ_i=even$,\ for these conifold-like singularities.
We find a cascade-like decay structure here too, including decays to 
lower order supersymmetric conifold-like singularities.

Finally we briefly discuss the physics of M2-branes stacked at
$\BC^4/\BZ_N$ singularities and nonsupersymmetric $AdS_4\times
S^7/\BZ_N$ backgrounds. The arguments of \cite{horopolchAdSInstab} for
a nonperturbative gravitational instability of nonsupersymmetric
$AdS_5\times S^5/\BZ_N$ backgrounds similar to the bubble-of-nothing
\cite{wittenBubble} decay of a Kaluza Klein vacuum apply to this case
also, suggesting a rapid decay of nonsupersymmetric $AdS_4\times
S^7/\BZ_N$ throat backgrounds.  However, it would seem that along the
lines of \cite{kachrutrivediAdSorbs}, cutting off the ultraviolet of
the throats, by \eg\ embedding in a compact space, would yield a finite 
decay rate. This then suggests the interesting possibility of stable 
nonsupersymmetric $AdS_4\times S^7/\BZ_N$ throat backgrounds in 
M-theory.

In sec.~2, we describe the twisted sector spectrum of
nonsupersymmetric $\BC^4/\BZ_N$ singularities, with a discussion of
terminal singularities in sec.~3, and of the phases of 4-dim orbifold
and conifold-like singularities in sec.~4. In sec.~5, we discuss
M2-branes stacked at $\BC^4/\BZ_N$ and the physics of
nonsupersymmetric $AdS_4\times S^7/\BZ_N$ backgrounds. Various 
details are contained in the appendices --- Appendix A describes 
aspects of the orbifold spectrum, Appendix B contains the Maple 
program we have used while Appendix C reviews aspects of GLSMs 
as applicable here, and Appendix D elucidates these GLSM techniques 
in some 4-dim supersymmetric singularities.

\section{$\BC^4/\BZ_N$ twisted sector spectrum}

The orbifold $\BC^4/\BZ_N (k_1,k_2,k_3,k_4)$ is defined by the 
action\ $z_i\ra e^{2\pi i jk_i/N} z_i,$\ where $z_i, i=1,\ldots,4$,\ 
are the four complexified coordinates in $\BC^4$, and $j=1,\ldots,N-1,$ 
labels the various twisted sectors. Such an orbifold with at least one 
of the $k_i$ coprime with $N$ satisfies 
$\left\{{j_0k_i\over N}\right\}={1\over N}$ for some sector $j_0$ and 
can be written in canonical form as $\BC^4/\BZ_N (1,p,q,r)$. These 
include all isolated orbifolds, with $p,q,r$ coprime with $N$. 
Orbifolds with no $k_i$ coprime w.r.t. $N$ are similar in structure 
to $\BC^3/\BZ_N$ orbifolds in the sector with some
$\{{j_0k_{i_0}\over N}\}=0$.

To understand the Type II GSO projection for nonsupersymmetric 
$\BC^4/\BZ_N (1,p,q,r)$ orbifolds, we complexify the fermions (before 
orbifolding) in each of the four physical 2-planes obtaining the spinor 
states\ 
$\{s_{ij}\}=\{\pm {1\over 2},\pm {1\over 2},\pm {1\over 2},\pm {1\over 2}\}
\equiv |\pm \pm \pm \pm \rangle$. The projection on these $SO(8)$ 
spinors to an irreducible chiral spinor requires\ $\sum s_{ij}=even$, 
restricting the spinor states to be 8-dimensional. 
Consider the (Green-Schwarz) orbifold rotation generator:
\be
\quad R={\rm exp}~[{2\pi i\over N} (J_{23}+pJ_{45}+qJ_{67}+rJ_{89})] .
\ee
Then\ $R^N=(-1)^{2(s_{23}+ps_{45}+qs_{67}+rs_{89})}$.\ For weights\ $(1,1,1,1)$, 
this gives\ $R^N=(-1)^{2(s_{23}+s_{45}+s_{67}+s_{89})}$. So clearly $R^N=1$ 
for any of the spinor states: this is always a Type II theory.\\
Now consider the spinor states $|\pm \pm \pm \pm \rangle$ that are 
invariant under the orbifold rotation generator $R$:\ 
\ $R |\pm\pm\pm\pm\rangle = (-1)^{2\sum s_{ij}\over N} |\pm\pm\pm\pm\rangle$ .
For $N=1$, the phase is $(-1)^{2\sum s_{ij}}$, which is trivial for all 
$2^4=16$ spinor states $s_{ij}$: so these preserve all (${\cal N}=8$) 
susy. For $N=2$, the phase is $(-1)^{\sum s_{ij}}$: this is trivial for 
states $|++++\rangle, |----\rangle$, $|++--\rangle$ and the 5 permutations 
thereof, which are precisely the 8 states from the chirality 
projection. Thus this also preserves\ ${\cal N}=8$ supersymmetry.
For $N\geq 3$, the phase is $(-1)^{(2\sum s_{ij})/N}$:\ thus only states 
with $|++--\rangle$ (and permutations) with $\sum s_{ij}=0$ give a 
trivial phase and are invariant. These are $6$ states, giving 
${\cal N}=6$ supersymmetry.

For a general $\BC^4/\BZ_N (1,p,q,r)$ orbifold, the rotation phase is
\be\label{Rphase}
R |\pm\pm\pm\pm\rangle = (-1)^{2 (s_{23}+ps_{45}+qs_{67}+rs_{89})\over N} 
|\pm\pm\pm\pm\rangle = (-1)^{(\pm 1\pm p\pm q\pm r)\over N} 
|\pm\pm\pm\pm\rangle .
\ee
So the phase is trivial if any of the combinations 
$\pm 1\pm p\pm q\pm r=0 (mod\ 2N)$. If \eg\ $1-p+q+r=0$, then there are 
two states $|+-++\rangle, |-+--\rangle$ with trivial phase. Thus if 
no such combination vanishes, then the orbifold completely breaks 
supersymmetry. This is the family of singularities we deal with in 
this paper. As we will see below, this dovetails well with the 
classification of twisted sector states into various chiral and 
anti-chiral rings.

From (\ref{Rphase}), we have\ 
$R^N |\pm\pm\pm\pm\rangle=(-1)^{(\pm 1\pm p\pm q\pm r)} |\pm\pm\pm\pm\rangle$. 
We require $R^N=1$ (rather than $R^N=(-1)^{\rm F}$, ${\rm F}$ being the 
spacetime fermion number), to remove the bulk tachyon retaining a Type 
II theory with spacetime fermions in the bulk, with possible closed 
string tachyons localized at the orbifold fixed point. A shift by an 
even integer $2p$ (or $2q,2r$) does not change the parity of a number, 
so this gives from the phase
\be\label{TypeIIGSO}
\pm 1\pm p\pm q\pm r=even \quad \Rightarrow\quad 1+p+q+r=even, 
\qquad \ie\qquad \sum_i k_i=even \ ,
\ee
for a $\BC^4/\BZ_N (k_1,k_2,k_3,k_4)$ orbifold, as the Type II GSO 
projection on the orbifold weights. A detailed RNS formulation of 
the Type II GSO projection appears in Appendix A.

The spectrum of twisted sector string excitations in a $\BC^4/\BZ_N
(k_1,k_2,k_3,k_4)$ orbifold conformal field theory, classified using
the representations of the $(2,2)$ superconformal algebra, has a
product-like structure (from each of the four complex planes): 
Appendix A describes various details, generalizing from the discussion 
of nonsupersymmetric $\BC^3/\BZ_N$ singularities \cite{drmknmrp}. Each 
complex plane contribution 
is either chiral ($c_{X_i}$) or antichiral ($a_{X_i}$), giving sixteen 
chiral and anti-chiral rings in eight conjugate pairs, labelled 
$(c_{X_1},c_{X_2},c_{X_3},c_{X_4}), (c_{X_1},c_{X_2},c_{X_3},a_{X_4})$, and 
so on.
These states can be succinctly described by the chiral ring twist field 
(vertex) operators, having the form $X_j=\prod_{i=1}^4 X^i_{\{jk_i/N\}}
=\prod_{i=1}^4\sigma_{\{jk_i/N\}}e^{i\{jk_i/N\}(H_i-{\bar H}_i)}$, 
where $\sigma_a$ is the bosonic twist-$a$ field operator, while the 
$H_i$ are bosonized fermions. These operators correspond to either 
the ground state or the first excited state in each twisted sector. 
For instance, in the sector where 
$\{{jk_1\over N}\},\{{jk_2\over N}\},\{{jk_3\over N}\}<{1\over 2} ,
\ \{ {jk_4\over N} \}>{1\over 2}$ , the ground state is of the form\ 
$\prod_{i=1}^3 X^i_{\{jk_i/N\}} (X^4_{1-\{jk_4/N\}})^*$, belonging to the 
$(c_{X_1},c_{X_2},c_{X_3},a_{X_4})$ ring (or simply $cccc$-ring). Then 
the $X_j$ are the first excited states in this sector, obtained by 
acting with\ $\psi^4\psi^{4*}=e^{i(H_i-{\bar H_i})}$ on the ground state 
operator. The conformal dimension of $X_j$ is\ 
$\Delta_j={1\over 2}\sum_i(\{{jk_i\over N}\}(1-\{{jk_i\over N}\})+{1\over 
2}(\{{jk_i\over N}\})^2)={1\over 2}\sum_i\{{jk_i\over N}\}$, and being 
chiral operators, they satisfy\ $\Delta_j={1\over 2} R_j$. 
The worldsheet R-charge $R_j$ and Type II GSO projection for the $X_j$ 
are\footnote{Note $\{x\}=x-[x]$ denotes the fractional part of $x$, 
with $[x]$ the integer part of $x$ (the greatest integer $\leq x$). 
By definition, $0\leq \{x\}<1$. Note that, for $m,n > 0$, we have 
$\ [{-m\over n}]=-[{m\over n}]-1\ $ and therefore 
$\{ {-m\over n} \}=-{m\over n}-[{-m\over n}]=1-\{ {m\over n} \}$.} 
\be
R_j\equiv \left(\left\{{jk_1\over N}\right\},\left\{{jk_2\over N}\right\},
\left\{{jk_3\over N}\right\},\left\{{jk_4\over N}\right\}\right)
=\sum_i \left\{{jk_i\over N}\right\}\ ,\ \quad\
E_j = \sum_i \left[{jk_i\over N}\right] = odd ,
\ee
\ie\ a GSO-allowed state has $X_j\ra (-1)^{E_j} X_j = -X_j$, the 
minus sign arising from the ghost contribution to the GSO exponent 
(in the $(-1,-1)$-picture) ensuring that the total worldsheet $(-1)^F$ 
is even for a GSO-preserved state.
The spacetime masses arising from the mass-shell condition is given by 
\be\label{massshell}
m_j^2={2\over\al'} (R_j-1)\ . 
\ee
The GSO exponent $E_j$ for a twist field operator $X_j$ depends 
nontrivially on the twist sector $j$ as well as the specific 
(anti-)chiral ring that the twist field belongs to. From the 
$\BC^4/\BZ_N\ (k_1,k_2,k_3,k_4)$ worldsheet partition function (see 
Appendix A), we obtain the GSO exponents (\ref{EjGSO}) for the sixteen 
rings. It is sufficient to discuss eight of these since the others just 
contain conjugate fields.

We now mention a convenient notation that can be used to study and label 
twist operators in the various rings. To illustrate this, note that 
twist operators in \eg\ the $(c_{X_1},c_{X_2},c_{X_3},a_{X_4})$-ring can 
be rewritten as\ 
$X_j^{ccca} = \prod_{i=1}^3 X^i_{\{jk_i/N\}} (X^4_{1-\{jk_4/N\}})^*
=\prod_{i=1}^3 X^i_{\{jk_i/N\}} (X^4_{\{-jk_4/N\}})^*$, 
which resemble twist operators in the $(c_{X_1},c_{X_2},c_{X_3},c_{X_4})$-ring 
of the orbifold $\BC^4/\BZ_N (k_1,k_2,k_3,-k_4)$ with $X^4\ra (X^4)^*$: 
the R-charges of the operators are identical while the condition on their 
GSO exponents $E_j=\sum_i [{jk_i\over N}]=even$\ (see Appendix A) 
is re-expressed as
\be
X^{ccca}_j=\prod_{i=1}^3 X^i_{\{jk_i/N\}} X^{4*}_{\{-jk_4/N\}}:\qquad 
E_j^{ccca}=\sum_{i=1}^3\left[{jk_i\over N}\right]+\left[-{jk_4\over N}\right]
=E_j+1+even=odd\ ,
\ee
so that as expected for a $cccc$-ring operator, the corresponding 
GSO exponent $E_j^{ccca}$ is odd. Generalizing, we see that operators in 
non-$cccc$-rings of the $\BC^4/\BZ_N\ (k_1,k_2,k_3,k_4)$ orbifold can be 
expressed as $cccc$-ring operators of a corresponding orbifold with 
related weights, with the GSO exponents appearing $uniformly$ $odd$ 
in this notation, \ie\ 
\bc
$X_j^r \ra (-1)^{E_j^r} X_j^r = - X_j^r$ \qquad GSO-allowed\ \ if \qquad
$E_j^r=odd$. 
\ec
This rewriting is particularly convenient in our discussion 
of all-ring terminality to follow.

It is important to label $\BC^4/\BZ_N (1,p,q,r)$ orbifolds appropriately 
in order to have a complete but also simple catalog. We will restrict 
$p,q,r>0$, defining thus the chiral ($cccc$) ring. As discussed above, 
the $ccca$-ring of this orbifold is then equivalent to the orbifold 
$\BC^4/\BZ_N (1,p,q,-r)$, in the sense that the twisted sector charges 
(and GSO projections) are the same, and similarly for the other six 
rings. In all, we then have
\bea
cccc\equiv (1,p,q,r) , \quad ccca\equiv (1,p,q,-r) , \quad
ccaa\equiv (1,p,-q,-r) , \quad caca\equiv (1,-p,q,-r) ,\nonumber\\
ccac\equiv (1,p,-q,r) , \quad cacc\equiv (1,-p,q,r) , \quad
caac\equiv (1,-p,-q,r) , \quad caaa\equiv (1,-p,-q,-r) .
\eea
Since we include all rings in listing the twisted sector spectrum, 
it is sufficient to restrict $0<p,q,r<N$:\ for instance, if $N<r<2N$, 
the orbifold $\BZ_N (1,p,q,r)\equiv \BZ_N (1,p,q,r-2N)$\ (shifting 
by $2N$ maintains the GSO projection), and the $cccc$-ring of 
the latter orbifold is equivalent to the $ccca$-ring of 
$\BZ_N (1,p,q,-(2N-r))$, which is contained within our restricted 
range since $0<2N-r<N$.

In this description, a supersymmetric orbifold is one where $some$ 
ring has a vanishing combination\ $1\pm p\pm q\pm r=0 (mod\ 2N)$, with 
GSO-preserved twisted states. For instance, with $1-p+q+r=0 (mod\ 2N)$, 
the $cacc$-ring is supersymmetric, with spectrum equivalent to\ 
$\BZ_N (1,-p,q,r)$. To illustrate this, consider $\BZ_N (1,p,q,r)$ with
say the $ccca$-ring being supersymmetric, \ie\ $1+p+q-r=0$. Then it can
be shown that no tachyons or moduli arise in any ring other than the 
$ccca$-ring. For instance, the $ccca$-ring R-charges are\
$R_j^{ccca}={j\over N}+\{{jp\over N}\}+\{{jq\over N}\}+1-\{{jr\over N}\}
={j(1+p+q-r)\over N}-E_j^{ccca}=-E_j^{ccca}$, for any $j$. Now if 
$R_j^{ccca}=1=-E_j^{ccca}$ for some sector $j$, then this is a twisted 
modulus. It is easily seen that the GSO exponents for the 
$cccc,ccaa$-rings (and permutations) are even, so these rings do not 
give any GSO-preserved states in such twist sectors. Also, using 
$R_j^{ccca}=1$, we have \eg\ 
$R_j^{cacc}=R_j^{ccca}+2(\{{jr\over N}\}-\{{jp\over N}\})>1$ and 
similarly, $R_j^{ccac},R_j^{caaa}>1$ (states in these rings are 
GSO-preserved). Now in sectors where $R_j^{ccca}=-E_j^{ccca}=2$, states 
in $cccc,ccaa$-rings (and permutations) are GSO-preserved: but we see
using $R_j^{ccca}=2$ that\ $R_j^{cccc}=1+2\{{jr\over N}\} ,\ 
R_j^{ccaa}=2+\{{-jq\over N}\}>1$, so that these states are irrelevant.
Thus only the supersymmetric ring (here $ccca$) contributes moduli.

The combinatorics of $\BC^3/\BZ_N$ is quite different \cite{drmknmrp} 
from $\BC^4/\BZ_N$, with no ``$\sum s_{ij}$ cancellation''.
To be specific, for $\BC^3/\BZ_N$, the weights $(1,1,1)$ do not yield 
a Type II GSO projection (as can be seen from (\ref{TypeIIGSO}), 
setting say $k_4=r=0$), so we must shift the weights to $(1,1,1-N)$, 
which admits a Type GSO projection for $N$ odd.
This latter orbifold is the simplest Type II analog of $\BZ_N (1,1,1)$, 
with R-charges:\\
$ccc$:\ $R_j=2\{{j\over N}\}+\{{j(1-N)\over N}\}=3{j\over N}, \ \ 
E_j=2[{j\over N}]+[{j(1-N)\over N}]=-j.$\\
$cca$:\ $R_j=2\{{j\over N}\}+\{{-j(1-N)\over N}\}=1+{j\over N}, \ \ 
E_j=2[{j\over N}]+[{j(N-1)\over N}]=j-1.$\\
$cac$:\ $R_j=\{{j\over N}\}+\{{-j\over N}\}+\{{j(1-N)\over N}\}
=1+{j\over N}, \ \ 
E_j=[{j\over N}]+[{-j\over N}]+[{j(1-N)\over N}]=-j-1.$\\
$caa$:\ $R_j=\{{j\over N}\}+\{{-j\over N}\}+\{{-j(1-N)\over N}\}
=2-{j\over N}, \ \ 
E_j=[{j\over N}]+[{-j\over N}]+[{j(N-1)\over N}]=j-2.$\\
So this always has a GSO-preserved tachyon in the $j=1$ sector for $N>3$\ 
(for $N=3$, this is the marginal blowup of the $\BZ_3 (1,1,-2)$ 
supersymmetric orbifold).\
For $\BC^2/\BZ_N (1,1)$, the R-charges are:\ \
$cc$:\ $R_j=2 {j\over N},\ \ E_j=2[{j\over N}]$ ,\ \
$ca$:\ $R_j=\{{j\over N}\}+\{{-j\over N}\}=1,\ \ 
E_j=[{-j\over N}]=-1.$\\ These are GSO-preserved moduli.

\section{Nonsupersymmetric terminal singularities}

The mass shell condition $m^2_j={2\over\al'} (R_j-1)$ above (see
(\ref{massshell})) shows that a twisted sector state with $R_j<1$ is
tachyonic ($m_j^2<0$), while one with $R_j=1$ is marginal. States with
$R_j>1$ are massive (irrelevant).  An orbifold is $terminal$ (or 
all-ring terminal) if $all$ twisted sector states from $all$ chiral 
and anti-chiral rings are irrelevant, \ie\ they all have\ $R_j>1$.
This means that the orbifold admits no geometric blowup modes and 
cannot be physically resolved by (relevant or marginal) worldsheet 
string modes.

We recall that for $\BC^2/\BZ_N$ nonsupersymmetric singularities 
\cite{hkmm}, the Hirzebruch-Jung minimal resolution ensures that a 
tachyon or modulus always arises in the chiral (or antichiral) ring, 
so that a $\BC^2/\BZ_N$ singularity is always resolved by twisted 
sector states in a single chiral (or antichiral) ring alone.

$\BC^3/\BZ_N$ singularities are more complicated: a single chiral (or
antichiral) ring might be terminal: these are ``geometric terminal
singularities'', comprising purely Kahler blowup modes, and are often 
referred to as terminal singularities in the mathematics literature.
From the physical point of view, we need to look at all the various
rings to understand if an orbifold is terminal, \ie\ both Kahler and
nonKahler blowup modes (or generic metric blowup modes). 

The Type II GSO projection introduces an additional complication by 
retaining only some states in each chiral ring, so that a possible 
geometric blowup mode could in fact be physically GSO-disallowed: 
this might suggest Type II terminal singularities are more likely.
However the combinatoric proof in \cite{drmknmrp} shows that a Type II
GSO-preserved tachyon or modulus always arises in one of the $j=1$
twisted sectors (all rings considered) for a $\BC^3/\BZ_N (1,p,q)$
orbifold in Type II string theories, while $\BC^3/\BZ_2 (1,1,1)$ is
the unique terminal singularity in Type 0 string theories.

For $\BC^4/\BZ_N$, it turns out that the $j=1$ twist sectors can be
terminal: we must analyze $all$ twisted sectors to see if tachyons or
moduli always arise. This makes the system $much$ more complicated and
a closed form proof to show the likely absence of terminal
singularities is difficult to obtain. However analyzing the $j=1$
sector gives various constraints on which $\BC^4/\BZ_N (1,p,q,r)$
singularities can be terminal if at all. It is then possible to
perform an ``experimental'' search using a Maple program (see Appendix
B) on these restricted window of possibilities for terminal
singularities. This reveals the absence of any nonsupersymmetric 
Type II terminal singularities as we run the Maple program through 
$N\leq 400$ for various orbifold weights, as we describe in greater 
detail below. There are of course Type 0 terminal singularities (as 
well as supersymmetric ones) as we will see below.

\subsection{Type II all-ring terminality}

It is straightforward to check that $\BC^4/\BZ_N (1,1,1,1)$ 
singularities are in fact terminal. Since the $U(4)$ symmetry is unbroken, 
we need to only check the twisted sector spectrum from the $cccc, ccca, 
ccaa, caaa$ rings (the others being permutations thereof).\\
$cccc$:\ $E_j=4[{j\over N}]=0\ \Rightarrow$ no GSO-preserved state.\\
$ccca$:\ 
$R_j=3\{{j\over N}\}+\{{j(-1)\over N}\}=3\{{j\over N}\}+1-\{{j\over N}\}
=1+2\{{j\over N}\}>1 \ \ (E_j=3[{j\over N}]+[{-j\over N}]=odd)$.\\
$caaa$:\ $R_j=\{{j\over N}\}+3\{{j(-1)\over N}\}=1+2(1-\{{j\over N}\})>1, 
\ \ (E_j=[{j\over N}]+3[{-j\over N}]=odd)$.\\ 
$ccaa$:\ $R_j=2\{{j\over N}\}+2\{{j(-1)\over N}\}=2,\ \ 
(E_j=2[{j\over N}]+2[{-j\over N}]=even)$.\\ 
Thus all GSO-preserved twisted sector states have R-charges $R_j>1$ and 
are irrelevant for all $N$: thus these are terminal singularities not 
resolvable by worldsheet blowups.

Note that there could be geometrically equivalent orbifolds admitting 
resolutions, essentially because they are different as conformal field 
theories due to a different GSO projection on the twisted states. For 
instance, $\BC^4/\BZ_4 (1,1,1,-3)$, although geometrically equivalent 
to $\BC^4/\BZ_4 (1,1,1,1)$ in fact has a marginal blowup mode arising 
in the $cccc$ ring $j=1$ sector: $R_{j=1}=1$, with 
$E_{j=1}=3[{j\over N}]+[{-3\over N}]=-1$. This is consistent with 
\cite{harvendra} which discusses a near-horizon supergravity solution 
for $M2$-branes stacked at a resolved $\BC^4/\BZ_4$ singularity.

Similarly $\BC^4/\BZ_N (1,1,p,p)$ singularities are 
terminal for $p$ coprime w.r.t. $N$: we see that \\
$cccc$:\ $E_j=2[{jp\over N}]=even.\quad ccaa$:\ $E_j=2[{-jp\over N}]=even. 
\Rightarrow$ no GSO-preserved state.\\
$ccca$:\ $R_j=2\{{j\over N}\}+\{{jp\over N}\}+\{-{jp\over N}\}>1.\qquad$
$cacc$:\ $R_j=\{{j\over N}\}+\{-{j\over N}\}+2\{{jp\over N}\}>1$.\\
$caca$:\ 
$R_j=\{{j\over N}\}+\{-{j\over N}\}+\{{jp\over N}\}+\{-{jp\over N}\}>1.\qquad$
$caaa$:\ $R_j=\{{j\over N}\}+\{-{j\over N}\}+2\{-{jp\over N}\}>1$.\\
Thus all twisted states that are possibly GSO-preserved are irrelevant. 
We mention that \cite{morrisonstevens} showed that geometric terminal 
$\BC^4/\BZ_N$ singularities must have weights of the form 
$\BZ_N (1,-1,p,-p)$, with $N,p$ coprime: this can be recognized as 
the $caca$-ring of the orbifolds in question here. These are in fact 
supersymmetric singularities as is well known. Note that orbifolds 
of this kind with $p$ not coprime w.r.t. $N$ do in fact contain moduli 
in their twisted spectrum. 

Now we come to nonsupersymmetric $\BC^4/\BZ_N (1,p,q,r)$ singularities.
Possible twisted sector string states that are GSO-preserved arise 
from any of the eight (pairs of) rings. In the $j=1$ sector, this 
requires $E_{j=1}=\sum_i [{k_i\over N}]=odd$: thus in $\BZ_N(1,p,q,r)$, 
with $p,q,r>0$, possible GSO-preserved states can only arise from the 
$ccca$, $ccac$, $cacc$, $caaa$ rings (for instance, 
$E_{j=1}^{ccaa}=[{1\over N}]+[{p\over N}]+[{-q\over N}]+[{-r\over N}]=2$). 
Terminality for these states gives the following conditions on their 
R-charges which must satisfy $R_j>1$:
\bea
{1\over N} + {p\over N} + {q\over N} + 1 - {r\over N} > 1\ ,\quad && 
{1\over N} + {p\over N} + 1 - {q\over N} + {r\over N} > 1\ ,\nonumber\\
{1\over N} + 1 - {p\over N} + {q\over N} + {r\over N} > 1\ ,\quad &&
{1\over N} + 1 - {p\over N} + 1 - {q\over N} + 1 - {r\over N} > 1\ ,
\eea
\be\label{ineq1}
\Rightarrow \quad 
1 + p + q > r\ , \quad 1 + p + r > q\ , \quad 1 + q + r > p\ , \quad 
p + q + r < 1 + 2N\ .
\ee
For Type II,\ \ $\sum k_i=even$, so that $p+q+r=odd$.\
Hence $r\neq p+q$ and similarly for $p,q$. Then the inequality\
$r>p+q \Longrightarrow\ p+q<r<p+q+1$ which is not possible for $r\in\BZ$. 
This has to hold for each of $p,q,r$: thus we must have (consistent 
if $p,q,r>0$)
\be\label{ineq1'}
r < p+q , \ \  q < p+r , \ \ p < q+r\ .
\ee
Furthermore, we must have\ $1\pm p\pm q\pm r\neq \nu N,\ \nu\in\BZ$,\ 
\ie\ the singularity is nonsusy in every ring. In particular,\ 
$1\pm p\pm q\pm r\neq 0$.\ This means
\be
1+p\gtrless q+r\ , \qquad 1+q\gtrless p+r\ , \qquad 1+r\gtrless p+q\ .
\ee
Now $1+r>p+q$\ means\ $p+q-1<r<p+q$, which is not possible for 
$r\in\BZ$, and similarly for the other inequalities. Thus we must have\
\be\label{ineq2}
r<p+q-1\ , \qquad q<p+r-1\ , \qquad p<q+r-1\ .
\ee
Combining (\ref{ineq1}), (\ref{ineq1'}) and (\ref{ineq2}) gives (recall 
$0<p<q<r$)
\be\label{ineq3}
q-p+1<r<p+q-1\ , \qquad r-p+1<q<p+r-1\ ,\qquad r-q+1<p<q+r-1\ ,
\ee
as the strongest inequalities.
Similar inequalities \eg\ $p-q-1<r<q-1+p$ are weaker: the lower bound 
is $p-q-1<0<r$ (since $p<q$).

If $p,q,r$, are not all distinct, we have a non-isolated singularity.\\

Say $p,q,r$, are all distinct: then we can take\ $0<p<q<r$ without 
loss of generality.
Now from the Type II condition $\sum_ik_i=$even, 
we must have (i) all odd, (ii) all even, (iii) 2 odd and 2 even. If 
we focus on isolated singularities, then $p,q,r,$ are mutually 
coprime, so that case (iii), with 2 even, is not allowed, nor is 
case (ii), all even. \\
Let us then restrict to $p,q,r$, all odd, and use the lowest such 
integers\  $p,q,r=1,3,5,7,9,11,13,\ldots$, restricting to mutually 
coprime integers for $(1,p,q,r)$.
\begin{itemize}
\item{$(1,p,q,r)$=(1,1,1,3), (1,1,3,5), (1,1,5,7), (1,1,7,9), (1,1,9,11), 
(1,1,11,13), (1,3,5,7),  (1,3,7,11), (1,5,7,11), (1,5,7,13), (1,5,9,13),
\ldots :\ these are supersymmetric, \eg\ the $ccca$-ring of 
$\BZ_N (1,1,5,7)$ is equivalent to the supersymmetric $\BZ_N (1,1,5,-7)$. }
\item{(1,1,1,5), (1,1,1,7), (1,1,1,11), (1,1,3,7), (1,1,5,9), (1,1,5,11), 
(1,1,5,13), (1,1,7,11), (1,1,7,13), (1,1,9,13), (1,3,7,13),\ldots :\ do 
not satisfy inequalities (\ref{ineq3}) above (in particular $r<p+q-1$). 
In other words, a tachyon arises in the $j=1$ sector in some ring.}
\item{(1,5,7,9), (1,7,9,11), (1,7,9,13), (1,5,11,13), (1,7,11,13),
 (1,9,11,13), (1,7,11,15) :\ satisfy (\ref{ineq3}) and potentially 
could be terminal. However the Maple program check for $N\leq 400$ 
shows no nonsupersymmetric terminal singularity\ (it does show 
supersymmetric terminal singularities \eg\ $\BZ_5 (1,7,9,13)\equiv 
\BZ_5 (1,2,-1,-2)$). More generally, we see that $(1,2m-1,2m+1,2m+3)$ 
and $(1,2m-3,2m+1,2m+5)$ satisfy (\ref{ineq3}) for $m>2$ and $m>4$ 
respectively, with $(1,3,5,7)$ and $(1,5,9,13)$ being supersymmetric.}
\item{Miscellaneous Maple checks for orbifolds with $N\leq 30$ and 
assorted weights show no terminal singularity.}
\end{itemize}


In general, the Maple output (see Appendix B) expectedly shows the
number of tachyons or moduli increasing as the orbifold order $N$
increases, thus making it less likely to find a terminal singularity
as $N$ increases. Indeed one expects to find a terminal singularity
for low orbifold orders, if at all: the absence thereof in the Maple
output is a noteworthy result.  Although our ``experimental'' search
is by no means exhaustive or equivalent to a closed form proof, not
finding any terminal singularity for the above checks and the
structure of the Maple output alongwith the analysis of the $j=1$
sector constraints above strongly suggests the non-existence of
nonsupersymmetric Type II terminal singularities.

By a close look at the Maple output, we find that 
\eg\ $\BZ_{13} (1,7,9,11)$ is supersymmetric, with $caaa$ ring, $j=1$ and 
$j=7$ twisted sector moduli:\ note\ $\{{-7.11\over 13}\}={1\over 13}$. 
Also\ $1-7-9-11=-26=-2N$ here, \ie\ saturation of the last inequality 
in (\ref{ineq3}). We also point out that \eg\ $\BZ_{17} (1,7,9,11)$ 
has (among others) a $cacc$ ring, $j=2$ twisted sector tachyon:\ 
note\ $\{{2.9\over 17}\}={1\over 17}$. 

In fact one might imagine this to be a general feature, \ie\ twisted 
sectors where\ $\{{\pm j_ak_b\over N}\}={1\over N}$\ for some sector 
$j=j_a$ and ring $k=\{k_b\}$ are likely to $always$ contain tachyons 
or moduli. While this is often true, it can be checked that the twisted 
sector tachyons of \eg\ the Type II orbifold $\BC^4/\BZ_{41} (1,7,9,11)$ 
do not arise from any such sector.

\subsection{Type 0 terminality}

The Type 0 theory has a diagonal GSO projection, the partition function 
being given in Appendix A. The spectrum can again be classified in 
terms of eight chiral and antichiral rings comprising operators $X_j$,
except that all such states exist and are GSO-preserved, \ie\ the GSO 
exponents are trivial.

For the Type 0 theory to be terminal, as a basic requirement, the 
$j=1$ sector should be terminal: this gives
\bea
{1\over N}+{p\over N}+{q\over N}+{r\over N}>1\ , \quad &&
{1\over N}+{p\over N}+{q\over N}+1-{r\over N}>1\ ,\nonumber\\
{1\over N}+{p\over N}+1-{q\over N}+{r\over N}>1\ , \quad &&
{1\over N}+1-{p\over N}+{q\over N}+{r\over N}>1\ ,\nonumber\\
{1\over N}+{p\over N}+1-{q\over N}+1-{r\over N}>1\ , \quad &&
{1\over N}+1-{p\over N}+{q\over N}+1-{r\over N}>1\ ,\nonumber\\
{1\over N}+1-{p\over N}+1-{q\over N}+{r\over N}>1\ , \quad &&
{1\over N}+1-{p\over N}+1-{q\over N}+1-{r\over N}>1\ ,
\eea
which simplify to
\bea
1+p+q+r>N\ , \quad p+q+1>r\ , \quad p+r+1>q\ , \quad q+r+1>p\ ,\nonumber\\
q+r<p+N\ , \quad p+r<q+N\ , \quad p+q<r+N\ , \quad p+q+r<1+2N\ ,
\eea
for Type 0 all-ring terminality in the $j=1$ sector.
Some of these can be combined and recast as
\be
N-1<p+q+r<2N+1\ ,\quad -3<p+q+r<3N\ .
\ee

It is possible to check that\ $\BC^4/\BZ_3 (1,1,1,2)$\ is an all-ring 
Type 0 terminal singularity: we have the R-charges
\bea
cccc: ({1\over 3},{1\over 3},{1\over 3},{2\over 3}),\quad
({2\over 3},{2\over 3},{2\over 3},{1\over 3}), \qquad
ccca: ({1\over 3},{1\over 3},{1\over 3},{1\over 3}),\quad
({2\over 3},{2\over 3},{2\over 3},{2\over 3}), \nonumber\\
ccac: ({1\over 3},{1\over 3},{2\over 3},{2\over 3}),\quad
({2\over 3},{2\over 3},{1\over 3},{1\over 3}), \qquad
ccaa: ({1\over 3},{1\over 3},{2\over 3},{1\over 3}),\quad
({2\over 3},{2\over 3},{1\over 3},{2\over 3}), \nonumber\\
caac: ({1\over 3},{2\over 3},{2\over 3},{2\over 3}),\quad
({2\over 3},{1\over 3},{1\over 3},{1\over 3}), \qquad
caaa: ({1\over 3},{2\over 3},{2\over 3},{1\over 3}),\quad
({2\over 3},{1\over 3},{1\over 3},{2\over 3}),
\eea
the other rings (cacc,caca) being permutations of these. These all clearly 
satisfy $R_j>1$, giving irrelevant states. \\
Similarly, $\BZ_4 (1,1,1,2), \BZ_4 (1,1,2,3), \BC^4/\BZ_5 (1,1,2,3)$, are 
also all-ring terminal singularities\footnote{
For instance, the $j=1$ sector R-charges for the various rings in 
$\BC^4/\BZ_5 (1,1,2,3)$ are 
\bea
cccc: ({1\over 5},{1\over 5},{2\over 5},{3\over 5}),
\quad \
ccca: ({1\over 5},{1\over 5},{2\over 5},{2\over 5}),
\ \ \ && 
ccac: ({1\over 5},{1\over 5},{3\over 5},{3\over 5}),
\quad \
cacc: ({1\over 5},{4\over 5},{2\over 5},{3\over 5}),
\ \ \nonumber\\
ccaa: ({1\over 5},{1\over 5},{3\over 5},{2\over 5}),
\quad \
caac: ({1\over 5},{4\over 5},{3\over 5},{3\over 5}),
\quad  && 
caca: ({1\over 5},{4\over 5},{2\over 5},{3\over 5}),
\quad \
caaa: ({1\over 5},{4\over 5},{3\over 5},{2\over 5}), 
\ \ ,\nonumber
\eea
the $j=2,3,4$ sectors being clearly irrelevant also.}.\ 
$\BZ_4 (1,2,3,5), \BZ_5 (1,3,4,7)$, can also be checked to be terminal, 
but can be recast as one of the above by shifting some of the weights 
(but retaining the Type 0 GSO projection).

The Maple output for Type 0 terminality in fact points out the above 
singularities but does not show any other. This is again of course 
not exhaustive by any means but suggests that as the orbifold order 
increases, Type 0 terminality does not occur either.


\section{$\BC^4/\BZ_N$ toric geometry, closed string tachyons 
and flips}

As we have seen, the twisted sector spectrum of nonsupersymmetric
$\BC^4/\BZ_N$ singularities shows localized closed string tachyonic
instabilities: the condensation of these tachyons causes a decay of
the system to more stable endpoints, which generically being unstable
also, decay. This process eventually stops when the system has no
further instabilities, \ie\ when all residual endpoints are either
fully smooth or supersymmetric singularities (which can be
terminal). Analysing the decay of such an unstable singularity is
elegantly done using gauged linear sigma models (GLSMs): a detailed
development of GLSMs for supersymmetric toric varieties was performed
in \cite{morrisonplesserInstantons}.  In the present nonsupersymmetric
context, they dovetail beautifully with the toric geometry description
of the resolution of these singularities (Appendix C reviews various
aspects of GLSMs applied to unstable noncompact singularities). These
GLSMs are in a sense simplified versions of nonlinear sigma models of
strings propagating on these unstable singularities (see \eg\
\cite{emilrev,minwalla0405} for reviews): localized closed string
tachyons are represented as relevant operators that induce
renormalization group flows from these unstable fixed points to more
stable fixed points typically representing lower order
singularities. The endpoints of the RG flows in the GLSM being
classical phases coincide with those of the nonlinear sigma model and
the GLSM RG flows themselves approximate the nonlinear ones in the
low-energy regime. The GLSMs here, which all have $(2,2)$ worldsheet
supersymmetry, have close connections with their topologically twisted
versions (the twisted A-models retain information about Kahler
deformations while complex structure information is in general lost):
thus various physical observables, in particular those preserving
worldsheet supersymmetry (\eg\ operators in chiral rings), are
protected along the RG flows corresponding to tachyon condensation
which give rise to Kahler deformations of the orbifold. Along the
flow, only part of the supersymmetry is preserved, that corresponding
to the chiral ring containing the condensing tachyon(s) of the parent
orbifold. However, at the end of the flow, the (more stable) fixed
points being residual orbifolds again have a twisted spectrum
comprising all their various chiral rings.

These worldsheet techniques were used to study the condensation of
closed string tachyons localized at lower dimensional ($\BC/\BZ_N$,
$\BC^2/\BZ_N$) orbifold singularities \cite{aps,vafa,hkmm}. They were
generalized to unstable orbifold $\BC^3/\BZ_N$ and conifold-like
$(\bA{cccc} n_1 & n_2 & -n_3 & -n_4 \eA)$ singularities in 3-complex
dimensions in \cite{drmknmrp,drmkn,knconiflips} (see also \eg\
\cite{sarkar0407}). The fact that single chiral or antichiral rings 
in 3-dim can be terminal makes the decay phase structure more
intricate. Since there are generically multiple decay channels
stemming from multiple tachyons, the most likely decay channel
corresponds to condensation of the most relevant tachyon (which
belongs in some ring), which induces a partial resolution of the
singularity: geometrically this is a weighted $\BC\BP^2$ expanding in
time.  Typically there are residual singularities on the expanding
locus which could be terminal with respect to the complex structure of
the ring containing the condensing tachyon. However since there are no
Type II terminal singularities, a tachyon (or modulus) in some other
ring will induce a blowup further resolving the system. Systems with
multiple tachyons generically exhibit flip transitions, \ie\ a
blowdown of a 2-cycle accompanied by a blowup of a topologically
distinct 2-cycle: in $\BC^3/\BZ_N$ orbifolds, this occurs when a more
dominant tachyon condenses during the condensation of some tachyon,
the tachyons being twisted sector states in the orbifold conformal
field theory.  In the context of conifold-like singularities, the 
instabilities do not appear to have a manifest conformal field theory 
interpretation, although the GLSM captures the geometric process 
adequately.

Our use of (first order) worldsheet RG flow to mimic (second order)
time evolution in spacetime is clearly an approximation: the RG time
of the GLSM agrees qualitatively with time in spacetime in known
examples, in the presence of worldsheet supersymmetry, for the special
kinds of complex noncompact singular spaces we deal with here. In \eg\
\cite{headrick0510, suyama}, it was found that for noncompact
singularities the worldsheet beta-function equations show no
obstruction to either RG flow (from c-theorems) or time-evolution
(since the dilaton can be turned off). Compact tachyons are more
intricate -- among other things, the dilaton is necessarily turned
on. We expect that an uplift to M-theory is consistent with this
structure of tachyon condensation and resolution of singularities,
using the metric variations and scalar duals (in 3-dims) to the $U(1)$
gauge fields (obtained from the 3-form C-field with some components on
the orbifold) as complex Kahler parameters entering in the geometric
(GLSM) description of the resolutions of the orbifold singularity.

\subsection{$\BC^4/\BZ_N$ toric geometry}

\begin{figure}
\bc
\epsfig{file=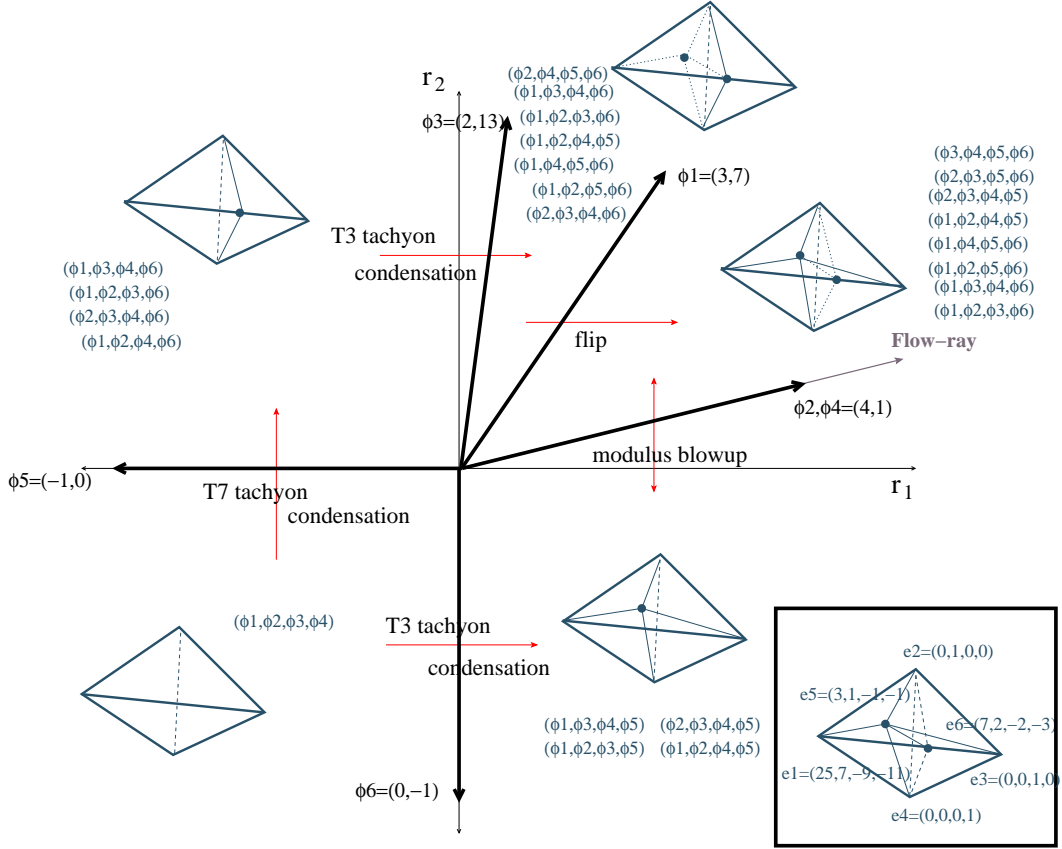, width=14cm}
\caption{{\small The 3-dim hyperplane of the\ $\BC^4/\BZ_{25} (1,-7,9,11)$ 
orbifold toric cone. The tachyonic lattice points (and the subdivisions 
thereof) depicted here are really the projections onto this 3-dim 
hyperplane of the actual points (which are in the 4-dim interior of 
the cone).}}
\label{figorb}
\ec
\end{figure}

The geometry of such an orbifold can be recovered efficiently using
its toric data. Let the toric cone $C(0;e_1,e_2,e_3,e_4)$ of this
orbifold be defined by the origin $0$ and lattice points
$e_1,e_2,e_3,e_4$ in the 4-dimensional toric $\BN$ lattice (the box in
Figure~\ref{figorb} shows the toric cone for $\BZ_{25} (1,-7,9,11)$):
the points $e_i$ define a 3-dimensional affine ``marginality
hyperplane'' (\ie\ tetrahedral cone) $\Delta$ passing through them. The
volume of this cone $V(0;e_1,e_2,e_3,e_4)\equiv |{\rm det}
(e_1,e_2,e_3,e_4)|$ gives the order $N$ of the orbifold singularity
(normalizing the cone volume without any additional numerical
factors). The specific structure of the orbifold represented by some
toric cone $C(0;e_1,e_2,e_3,e_4)$ can be gleaned either using the
Smith normal form algorithm \cite{drmknmrp}, or equivalently by
realizing relations between the lattice vectors $e_i$ and any vector
that is also itself contained in the toric $\BN$ lattice: \eg\ we see
that the cone defined by\ $e_1=(N,-p,-q,-r), e_2=(0,1,0,0),
e_3=(0,0,1,0), e_3=(0,0,0,1)$, corresponds to $\BC^4/\BZ_N (1,p,q,r)$
using the relation $(1,0,0,0)={1\over N}(e_1+pe_2+qe_3+re_4)$ with the
lattice point $(1,0,0,0)$. Note that in general this only fixes the
orbifold weights upto shifts by the order $N$.

A $\BC^4/\BZ_N (1,p,q,r)$ orbifold is isolated if $p,q,r$ are coprime
w.r.t. $N$: this is equivalent to the condition that there are no
lattice points on the walls of the defining toric cone. For example,
if $q,N$ have a common factor $n$ with $q=m_1n,\ N=m_0n$, then the
$\{e_1,e_2\}$ and $\{e_1,e_4\}$ walls have the integral lattice points
${1\over n}(N,-p,-q,-r)+\{{p\over n}\}(0,1,0,0)=(m_0,-[{p\over n}],-m_1,0)$ 
and ${1\over n}(N,-p,-q,-r)+\{{r\over n}\}(0,0,0,1)=(m_0,0,-m_1,-[{r\over N}])$
respectively.

Geometric terminal singularities arise if there is no K\"ahler blowup
mode: \ie\ no lattice point in the interior of the toric cone and
equivalently no relevant or marginal chiral ring operator in the
orbifold spectrum. A physical analysis of the system must include all
possible tachyons in all rings, \ie\ both K\"ahler and non-K\"ahler
blowup modes: this dovetails with our discussion above on the twisted
spectrum and the absence of terminal singularities. Note also that
$\BC^4/\BZ_N$ singularities (as in $\BC^3/\BZ_N$) have no complex
structure deformations \cite{schless}: generically these are
incomplete intersections.

There is a 1-1 correspondence between the chiral ring operators and 
points in the $\BN$ lattice toric cone of the orbifold. A given lattice 
point $P_j=(x_j,y_j,z_j,w_j)$ can be mapped to a twisted sector chiral ring
operator in the orbifold conformal field theory by realizing that this 
vector can expressed in the $ \{e_1,e_2,e_3,e_4\} $ basis as
\be\label{Rjlattpt}
(x_j,y_j,z_j,w_j) = r_1e_1 + r_2e_2 + r_3e_3 + r_4e_4\ . 
\ee
If $0<r_i\leq 1$, then $P_j$ is in the interior of the cone, and 
corresponds to an operator $O_j$ with R-charge\ $R_j\equiv(r_1,r_2,r_3,r_4)$. 
Conversely, it is possible to map an operator $O_j$ of given R-charge 
to a lattice point $P_j$. Concretely, we can check using (\ref{Rjlattpt}) 
that a $cccc$-ring operator $X_j$ in twist sector-$j$ of 
$\BC^4/\BZ_N (1,p,q,r)$ with R-charge $R_j$ corresponds to a lattice 
point as
\be\label{RjPj}
P_j=\left(j,-\left[{jp\over N}\right],
-\left[{jq\over N}\right],-\left[{jr\over N}\right]\right)
\quad \equiv \quad R_j=\left({j\over N},\left\{{jp\over N}\right\},
\left\{{jq\over N}\right\},\left\{{jr\over N}\right\}\right) .
\ee
There are always lattice points lying ``above'' (in the 4-dimensional 
sense) the ``marginality hyperplane'' $\Delta$, corresponding to 
irrelevant operators: these have $R_j=\sum_ir_i>1$. Interior points 
lying $on$ $\Delta$ (\ie\ within the tetrahedral cone) have $R_j=1$ and 
are marginal operators (moduli), while those ``below'' (in 4D) the 
hyperplane $\Delta$ have $R_j<1$ and correspond to tachyons\footnote{Note 
that for the $\BC^4/\BZ_N (1,p,q,r)$ orbifold, we have the relation\
${x_j(1+p+q+r)\over N} + y_j + z_j + w_j = r_1+r_2+r_3+r_4=R_j ,$
so that for a supersymmetric orbifold $1+p+q+r=0 (mod\ 2N)$, we 
have all $R_j$ integral since $x_j,y_j,z_j,w_j\in\BZ$, \ie\ there are no 
tachyonic lattice points.}. Roughly, the more relevant a tachyon, 
\ie\ the smaller $R_j$, the deeper its lattice point is in the 
interior of the cone.
This orbifold toric cone can be subdivided by any of the 
tachyonic or marginal blowup modes: the irrelevant ones are unimportant 
physically (see \eg\ \cite{aspinwallOrbres}, which reviews such 
toric subdivisions). Heuristically, since the tachyon 
is a lattice point in the interior of the cone, a subdivision means 
removing the ``top'' subcone $C(T;e_1,e_2,e_3,e_4)$, retaining the 
four residual subcones $C(0;e_1,e_2,e_3,T), C(0;e_1,e_2,e_4,T), 
C(0;e_1,e_3,e_4,T), C(0;e_2,e_3,e_4,T)$. These are potentially 
orbifold singularities again, unstable to tachyon condensation. The 
cumulative volume of the four subcones obtained from a subdivision 
induced by a lattice point corresponding to a twisted sector operator 
of R-charge $R_j$ is $NR_j$: for a tachyon $R_j<1$, this cumulative 
volume, representing the cumulative order of the residual 
singularities, is less than the original orbifold order $N$, 
indicating a partial resolution of the singularity.
For example, condensation of the tachyon 
$T$ with $R_T\equiv ({1\over N},{p\over N},{q\over N},{r\over N})$ in 
the $\BC^4/\BZ_N (1,p,q,r)$ orbifold, corresponds to the subdivision of 
the cone $C(0;e_1,e_2,e_3,e_4)$ by the interior lattice point 
$T\equiv(1,0,0,0)$. From the GLSM point of view, this corresponds to 
RG flow of the single Fayet-Iliopoulos parameter in a GLSM with a 
$U(1)$ gauge group and 
charge matrix $Q=(\bA{ccccc} 1 & p & q & r & -N \eA)$: this gives the 
resolved phase as the stable phase. With the $U(1)$ action on 
$\Psi_i\equiv (\phi_1,\phi_2,\phi_3,\phi_4,T)$ being
$\Psi_i\ra e^{2\pi iQ_i\lambda} \Psi_i$, the D-term equation is 
(equivalently by the symplectic quotient construction)
\bea
-D\equiv |\phi_1|^2 + p|\phi_1|^2 + q|\phi_1|^2 + r|\phi_1|^2 - N|T|^2 
= r//U(1)\ , \nonumber
\eea
the 1-loop renormalization of $r$ being 
$r={(1+p+q+r-N)\over 2\pi} \log {\mu\over\Lambda}$ . The coefficient 
is $N (R_T-1)<0$, so that $r$ flows from $r<0$ in the ultraviolet 
$\mu\gg\Lambda$ to $r>0$ in the infrared $\mu\ll\Lambda$. For $r<0$, 
$T$ must have a nonzero expectation value, which with the action\
$T\ra e^{-2\pi iN\lambda} T$ , Higgses the $U(1)$ down to a residual 
$\BZ_N$ acting on the light fields $\phi_i$ as 
$\phi_i\ra e^{2\pi iQ_i {j\over N}} \phi_i$, giving the orbifold 
$\BZ_N (1,p,q,r)$. Alternatively for $r>0$, one of the $\phi_i$ must 
acquire an expectation value, leaving the light fields 
$\{\phi_1,\phi_2,\phi_3,T\}$ and other permutations, which give the 
coordinate charts describing the blown-up $w\BC\BP^3$ (with residual 
$\BZ_p,\BZ_q,\BZ_r$, orbifold singularities). The partial resolution 
in this case typically occurs by the blowup of a weighted $\BC\BP^3$ 
with potentially four residual orbifold singularities, as the GLSM 
D-term shows\footnote{See \eg\ \cite{0308028}, which uses the mirror 
Landau-Ginzburg description of \cite{vafa} to show that under 
condensation of a single tachyon, a $\BC^r/\BZ_N$ orbifold decays 
into $r$ separated orbifolds.}. From the holomorphic quotient point 
of view, introduce coordinates $x_i, i=1,\ldots,5$, corresponding to
the lattice points $e_i,T$, with the $\BC^*$ quotient action 
$x_i\ra\lambda^{Q_i}x_i$ with $\lambda\in\BC^*$. Then the divisors 
(complex codim one hypersurfaces) $x_i=0, i=1,\ldots,4$ are 
noncompact, while $x_5=0$ is a compact divisor: on $x_5=0$, the 
$\BC^*$ action is\ $(x_1,x_2,x_3,x_4,0) \sim\ 
(\lambda x_1, \lambda^p x_2, \lambda^q x_3, \lambda^r x_4, 0)$. For 
a finite size divisor, we expect a non-degenerate description of 
the 4-dim space: we must therefore exclude the set 
$(x_1,x_2,x_3,x_4)=(0,0,0,0)$. This yields the weighted projective 
space $\BC\BP^3_{1,p,q,r}$, described by the coordinate chart 
$(x_1,x_2,x_3,x_4)$, equivalent to the symplectic quotient we 
use here. 
Systems of multiple tachyons in orbifolds can be analyzed by
appropriate generalizations of this GLSM as for $\BC^3/\BZ_N$
orbifolds \cite{drmkn} (Appendix C reviews aspects of GLSMs in this
context), and generically exhibit 4-dimensional flip transitions 
amidst their phases as we will see below.

As is usually the case, the dimensions (or R-charges) of various
operators are renormalized under an RG flow induced by some relevant
operator (say tachyon $T_1$). An interesting feature of these
worldsheet supersymmetric systems is that the R-charges of residual
tachyons can be calculated using the combinatorics of the toric fan
(as for $\BC^3/\BZ_N$, where visualization of the toric cone, being 
3-dim, was easier!). Since a residual tachyon $T_2$ is contained in 
the interior (or the ``walls'') of a subcone say
$C(0;e_1,e_2,e_3,T_1)\equiv\BZ_{N'}$, one of whose defining lattice
points is the tachyon $T_1$, we have a relation of the form\
$T_2={1\over N'} (r_1e_1+r_2e_2+r_3e_3+r_4T_4)$ with $N'<N$. Since the
marginality hyperplane $\Delta'$ of this subcone dips inward relative
to $\Delta$ of the original orbifold, the residual tachyon $T_2$ is
closer to $\Delta'$ than it was to $\Delta$. This means $T_2$ must
become more irrelevant under the RG flow of $T_1$.  This is a fairly
general statement: a tachyon always becomes less tachyonic (\ie\ more
irrelevant) under condensation of some tachyon, a fact that can be
checked explicitly and is in fact borne out in the examples below.

There are also important consequences of the GSO projection for the
residual orbifold subcones and the lattice points in their
interior. Recall that an orbifold $\BC^4/\BZ_N (k_1,k_2,k_3,k_4)$
admits a Type II GSO projection if $\sum_ik_i=even$ and a twist
sector-$j$ operator $X_j$ with R-charge $R_j$ is GSO-preserved if
$E_j=\sum_i \left[{jk_i\over N}\right]=odd$. It can be shown that
under condensation of a GSO-preserved tachyon 
$T_j=(j,-[{jp\over N}],-[{jq\over N}],-[{jr\over N}])\equiv
({j\over N},\{{jp\over N}\},\{{jq\over N}\},\{{jr\over N}\})$, the 
GSO projection for the residual orbifolds and residual tachyons is
consistent with this description, \ie\ each of the four residual 
orbifolds admits a Type II GSO projection. To show this, recall 
that the Type II GSO projection requires that $p+q+r=odd$ and 
$[{jp\over N}]+[{jq\over N}]+[{jr\over N}]=odd$. The four 
resulting subcones $C(0;T_j,e_2,e_3,e_4), C(0;T_j,e_1,e_2,e_4), 
C(0;T_j,e_1,e_3,e_4), C(0;T_j,e_1,e_2,e_3)$, are orbifolds 
$\BC^4/\BZ_n\ (w_1,w_2,w_3,w_4)$, with some weights $w_i$. The 
subcone $C(0;T_j,e_2,e_3,e_4)$, with the defining lattice points 
being in canonical form, can be recognized as 
$\BC^4/\BZ_j (1,[{jp\over N}],[{jq\over N}],[{jr\over N}])$, 
which is manifestly Type II. The lattice relation 
\bc
{$(1,0,0,0)={1\over N\{{jp\over N}\}} (pT_j - [{jp\over N}] e_1 
+ (p[{jq\over N}] - q[{jp\over N}]) e_3 
+ (p[{jr\over N}] - r[{jp\over N}]) e_4) $}\ec  
shows that the subcone $C(0;T_j,e_1,e_3,e_4)$ is the orbifold\\ 
$\BC^4/\BZ_{N\{{jp\over N}\}} (p,-[{jp\over N}],
p[{jq\over N}]-q[{jp\over N}],p[{jr\over N}]-r[{jp\over N}]),$ \
the orbifold action being on the coordinates represented by 
$T_j,e_1,e_3,e_4$ respectively. Such a linear combination of 
lattice vectors giving a vector in the original lattice is only 
defined up to adding integer multiples of the lattice vectors.
(If any of the coefficients of $T_j,e_1,e_3,e_4$ vanish, the subcone 
corresponds to a non-isolated, or lower-dim orbifold.) Now from 
the weights, we see that
\bc
$p-[{jp\over N}]+p[{jq\over N}]-q[{jp\over N}]+p[{jr\over N}]-r[{jp\over N}]
= p (1+[{jp\over N}]+[{jq\over N}]+[{jr\over N}]) - (1+p+q+r) [{jp\over N}]
= even$,
\ec
\ie\ the subcone $C(0;T_j,e_1,e_3,e_4)$ is a Type II orbifold. 
Similarly the other subcones can be shown to be Type II orbifolds.
Also it can be shown that originally GSO-preserved residual tachyons 
continue to be GSO-preserved after condensation of a GSO-preserved 
tachyon for each of the four residual singularities.

While the construction of the toric fan is a straightforward
generalization from that of\ $\BC^3/\BZ_N$, visualization is now 
difficult, especially when trying to understand residual tachyons or
moduli within a particular subcone obtained by some tachyon or modulus. 
Algorithmically then, it is more convenient to find the twisted sector
spectrum of an orbifold, note the most relevant tachyons arising
therein, and then analyse the phases of the corresponding gauged
linear sigma model (GLSM) to glean the structure of flips and the
dynamics of these orbifolds.

For example, in $\BZ_{19} (1,5,7,9)$, the most relevant (GSO preserved) 
tachyon $T_8$ lies in the $ccaa$-ring, the next most relevant tachyons 
arising in several distinct rings. There are two GSO preserved tachyons 
$T_8,T_4$ of R-charges 
$R_8\equiv ({8\over 19},{2\over 19},{1\over 19},{4\over 19})={15\over 19}$ 
and $R_4\equiv ({4\over 19},{1\over 19},{10\over 19},{2\over 19})={17\over 19}$ 
in the $ccaa$-ring (with spectrum equivalent to the $cccc$-ring of 
$\BZ_{19} (1,5,-7,-9)$, the R-charges being\ $R_j\equiv 
(\{{j\over 19}\},\{{5j\over 19}\},\{{-7j\over 19}\},\{{-9j\over 19}\})$):\ 
the GSO exponents 
$E_j^{ccaa}=[{j\over 19}]+[{5j\over 19}]+[-{7j\over 19}]+[-{9j\over 19}]$ 
can be checked to be $odd$, thus preserving the tachyons. The phase 
structure of the geometry and its blowups induced by the condensation 
of these tachyons can be analysed by a GLSM with charge matrix
\be\label{Qia191579}
Q_i^a = \left( \bA{cccccc} 4 & 1 & 10 & 2 & -19 & 0  \\ 8 & 2 & 1 
& 4 & 0 & -19  \\ \eA \right)\ .
\ee
The phase boundaries are represented by the rays\ $\phi_1, 
\phi_2,\phi_4\equiv (1,2),\ \phi_3\equiv (10,1),\ \phi_5\equiv (-19,0),\ 
\phi_6\equiv (0,-19)$. The flow-ray is the vector $F\equiv (1,2)$.
The relations $T_4={1\over 2} (T_8+e_3)$ and 
$T_8={1\over 10} (T_4+4e_1+e_2+2e_4)$ show that the $T_4$ lattice point 
lies on the $T_8,e_3$-wall and is collinear with $T_8,e_3$, while $T_8$ 
lies in the interior of the subcone $C(0;e_1,e_2,e_4,T_4)$. They also 
show that $T_4$ becomes marginal after condensation of $T_8$, while 
$T_8$ acquires the renormalized R-charge $R_8'\equiv 
({1\over 10},{4\over 10},{1\over 10},{2\over 10})={4\over 5}$.
Analyzing the coordinate charts in the phase diagram shows the four 
phases to correspond to the unresolved orbifold, the partial blowups 
induced by condensation of $T_4$ or $T_8$ and the complete blowup induced 
by condensation of both tachyons $T_8,T_4$. The stable phases correspond 
to the condensation of $T_8$ alone and of that of $T_8,T_4$. The details 
can be worked out using the techniques that we will describe below.
\\

\noindent {\bf Flip transitions:}\\
In more interesting cases, there are 4-dimensional flip transitions 
\cite{drmkn}: these occur when a more relevant tachyon condenses during 
condensation of some tachyon, causing a transition between two 
topologically distinct resolution endpoints. For instance, in 
$\BZ_{25} (1,7,9,11)$, the most relevant tachyon $T_3$ with R-charge 
$R_3\equiv ({3\over 25},{4\over 25},{2\over 25},{8\over 25})={17\over 25}$ 
lies in the $cacc$-ring: this has spectrum equivalent to the $cccc$-ring 
of $\BZ_{25} (1,-7,9,11)$, so that we can, if we wish, effectively define 
the orbifold in question here as\ $\BC^4/\BZ_{25} (1,-7,9,11)$. There are 
two more GSO preserved tachyons $T_7,T_{14}$, in this ring, $T_{14}$ being 
more relevant with R-charge $R_{14}={21\over 25}$ . However the structure 
of decay of the orbifold induced by $T_3,T_7$ exhibits more features so 
we focus on these in what follows. The R-charges for this ring are\ 
$R_j\equiv 
(\{{j\over 25}\},\{{-7j\over 25}\},\{{9j\over 25}\},\{{11j\over 25}\})$), 
and the GSO exponents 
$E_j^{ccaa}=[{j\over 25}]+[-{7j\over 25}]+[{9j\over 25}]+[{11j\over 25}]$ 
can be checked to be $odd$, thus preserving the tachyons. The tachyon 
$T_7$ has R-charge 
$R_7\equiv ({7\over 25},{1\over 25},{13\over 25},{2\over 25})={23\over 19}$ .

\noindent The phase structure of the geometry and its blowups induced by 
the condensation of these tachyons can be analysed by a GLSM with charge 
matrix\footnote{Including all three tachyons can be analysed by a 
3-parameter GLSM with charge matrix
\bea
Q_i^a = \left( \bA{ccccccc} 3 & 4 & 2 & 8 & -25 & 0 & 0 \\ 7 & 1 & 13 
& 2 & 0 & -25 & 0 \\ 14 & 2 & 1 & 4 & 0 & 0 &-25 \eA \right)\ . \nonumber
\eea
The flow-ray for this system is $(4,1,2)\equiv\phi_2$. It is possible 
to analyse this system using the secondary fan and find the stable 
phases.}
\be\label{Qia2517911}
Q_i^a = \left( \bA{cccccc} 3 & 4 & 2 & 8 & -25 & 0  \\ 7 & 1 & 13 
& 2 & 0 & -25  \\ \eA \right)\ .
\ee
The phase boundaries are represented by the rays\ $\phi_1\equiv (3,7), 
\phi_2,\phi_4\equiv (4,1),\ \phi_3\equiv (2,13),\ \phi_5\equiv (-25,0),\ 
\phi_6\equiv (0,-25)$. The flow-ray is the vector $F\equiv (4,1)$.
The cone is defined by the bounding vectors\ $(25,7,-9,-11), (0,1,0,0), 
(0,0,1,0), (0,0,0,1)$, with the tachyon lattice points being\ 
$T_3\equiv (3,1,-1,-1),\ T_7\equiv (7,2,-2,-3)$. We have the relations\
$T_3={3\over 25}e_1+{4\over 25}e_2+{2\over 25}e_3+{8\over 25}e_4$ and\ 
$T_7={7\over 25}e_1+{1\over 25}e_2+{13\over 25}e_3+{2\over 25}e_4$.
The relations\ $T_7={1\over 4}e_1+{1\over 2}e_3+{1\over 4}T_3$\ and\
$T_3={1\over 13}e_1+{2\over 13}e_2+{4\over 13}e_4+{2\over 13}T_7$\ 
respectively show that the $T_7$ lattice point lies on the plane 
containing $e_1,e_3,T_3$ (rather than in the interior of any subcone 
defined by $T_3$ with some three of the four points $e_i$), while the 
$T_3$ lattice point lies in the interior of the subcone 
$C(0;e_1,e_2,e_4,T_7)$. The $T_7$ relation also shows (using 
(\ref{Rjlattpt})) that after condensation of $T_3$, the tachyon $T_7$ 
acquires a renormalized R-charge $R_7'=1$, thus becoming marginal. 

The D-term conditions (alternatively the symplectic quotient) are
\bea
&& -D_1\equiv -D_{\phi_6}\equiv 
3|\phi_1|^2+4|\phi_2|^2+2|\phi_3|^2+8|\phi_4|^2-25|T_3|^2=r_1\ ,
\nonumber\\
&& -D_2\equiv -D_{\phi_5}\equiv 
7|\phi_1|^2+|\phi_2|^2+13|\phi_3|^2+2|\phi_4|^2-25|T_7|^2=r_2\ ,
\eea
with $r_1,r_2$ being the two Fayet-Iliopoulos parameters representing 
closed string blowup modes. These have the 1-loop renormalizations\ 
$r_1=({-8\over 2\pi})\log {\mu\over\Lambda}$ and 
$r_2=({-2\over 2\pi})\log {\mu\over\Lambda}$. By eliminating the 
appropriate coordinate field, we obtain the auxiliary D-terms useful 
for gleaning properties of the system crossing phase boundaries:
\bea
&& -{D_{\phi_1}\over 25} \equiv 
|\phi_2|^2-|\phi_3|^2+2|\phi_4|^2-7|T_3|^2+3|T_7|^2={7r_1-3r_2\over 25}\ ,
\nonumber\\
&& -{D_{\phi_2}\over 25} = -{D_{\phi_4}\over 25} \equiv 
-|\phi_1|^2-2|\phi_3|^2-|T_3|^2+4|T_7|^2={r_1-4r_2\over 25}\ ,\\
&& -{D_{\phi_3}\over 25} \equiv 
|\phi_1|^2+|\phi_2|^2+4|\phi_3|^2-13|T_3|^2+2|T_7|^2={13r_1-2r_2\over 25}\ .
\nonumber
\eea
Using these D-term equations and the renormalization group flowlines, we 
can realize the phase structure of this unstable orbifold (see the phase 
diagram (Fig~\ref{figorb})). For instance, in the convex hull 
$\{\phi_2,\phi_6\}\equiv \{\phi_4,\phi_6\}$, with $r_1>0,\ r_1>4r_2$, 
the D-terms $D_{\phi_6},D_{\phi_2}\equiv D_{\phi_4}$, imply that at least 
one element of each set $\phi_1,\phi_2,\phi_3,\phi_4$, and $T_7$ must 
acquire nonzero vacuum expectation values: the D-term equations do 
not have solutions for all of these simultaneously zero (this is 
the excluded set in this phase).
Now in the region of moduli space where $\phi_1,T_7$, acquire vevs, 
the light fields at low energies are $\phi_2,\phi_3,\phi_4,T_3$, which 
yield a description of the coordinate chart $(\phi_2,\phi_3,\phi_4,\phi_5)$. 
If $\phi_2,T_7$, acquire vevs, the light fields describe the chart 
$(\phi_1,\phi_3,\phi_4,\phi_5)$. Similarly we also obtain the charts 
$(\phi_1,\phi_2,\phi_4,\phi_5)$ and $(\phi_1,\phi_2,\phi_3,\phi_5)$ 
if $\phi_3,T_7$, and $\phi_4,T_7$, acquire vevs respectively. Note 
that each of these collections of nonzero vevs are also consistent 
with the other D-terms. Similarly we can analyse the other convex 
hulls. A simple operational method \cite{drmkn} to realize the
results of the above analysis of the D-terms for the phase boundaries
and the GLSM phases is the following: read off each column in
$Q_i^a$ given in (\ref{Qia2517911}) as a ray drawn out from the origin
$(0,0)$ in $(r_1,r_2)$-space, representing a phase boundary. Then 
the various phases are given by the convex hulls\footnote{A 
2-dimensional convex hull is the interior of a region bounded by two 
rays emanating out from the origin such that the angle subtended by 
them is less than $\pi$.} bounded by any two of the five phase 
boundaries represented by the rays $\phi_1\equiv (3,7),\ 
\phi_2\equiv\phi_4\equiv (4,1),\ \phi_3\equiv (2,13),\ 
\phi_5\equiv (-1,0),\ \phi_6\equiv (0,-1)$. These phase boundaries 
divide $r$-space into five phase regions, each described, as a 
convex hull of two phase boundaries, by several possible overlapping 
coordinate charts obtained by noting all the possible convex hulls 
that contain it. For instance, the coordinate charts describing the 
convex hull $\{\phi_2,\phi_6\}\equiv \{\phi_4,\phi_6\}$ are read off 
as the complementary sets $\{\phi_1,\phi_3,\phi_4,\phi_5\},\ 
\{\phi_1,\phi_2,\phi_3,\phi_5\}$. This convex hull is contained in 
the convex hulls $\{\phi_1,\phi_6\},\ \{\phi_3,\phi_6\}$: thus 
the full set of coordinate charts characterizing the toric variety 
in the phase given by the convex hull $\{\phi_2,\phi_6\}\equiv 
\{\phi_4,\phi_6\}$ is indeed what we have obtained above using 
the D-term equations.

The coordinate charts describing the phases of this orbifold, obtained 
as above, are
{\small 
\bea
\{\phi_5,\phi_6\}:\ (\phi_1,\phi_2,\phi_3,\phi_4), && 
\{\phi_3,\phi_5\}:\ (\phi_1,\phi_3,\phi_4,\phi_6),\ 
(\phi_1,\phi_2,\phi_3,\phi_6),\ (\phi_2,\phi_3,\phi_4,\phi_6),\ 
(\phi_1,\phi_2,\phi_4,\phi_6),\nonumber\\
\{\phi_2,\phi_6\}\equiv \{\phi_4,\phi_6\}: &&  (\phi_1,\phi_3,\phi_4,\phi_5),\ 
(\phi_1,\phi_2,\phi_3,\phi_5),\ (\phi_2,\phi_3,\phi_4,\phi_5),\ 
(\phi_1,\phi_2,\phi_4,\phi_5),\nonumber\\
\{\phi_1,\phi_2\}\equiv \{\phi_1,\phi_4\}: && (\phi_3,\phi_4,\phi_5,\phi_6),\ 
(\phi_2,\phi_3,\phi_5,\phi_6),\ (\phi_2,\phi_3,\phi_4,\phi_5),\ 
(\phi_1,\phi_2,\phi_4,\phi_5),\nonumber\\ 
&& (\phi_1,\phi_4,\phi_5,\phi_6),\ (\phi_1,\phi_2,\phi_5,\phi_6),\ 
(\phi_1,\phi_3,\phi_4,\phi_6),\ (\phi_1,\phi_2,\phi_3,\phi_6),\nonumber\\
\{\phi_1,\phi_3\}: && (\phi_1,\phi_3,\phi_4,\phi_6),\ 
(\phi_1,\phi_2,\phi_3,\phi_6),\ (\phi_2,\phi_3,\phi_4,\phi_6),\ 
(\phi_2,\phi_4,\phi_5,\phi_6), \nonumber\\ 
&& (\phi_1,\phi_2,\phi_4,\phi_5),\ (\phi_1,\phi_4,\phi_5,\phi_6),\ 
(\phi_1,\phi_2,\phi_5,\phi_6) . \nonumber
\eea
}
This shows that the convex hull $\{\phi_5,\phi_6\}$ is the unresolved 
orbifold phase, while $\{\phi_3,\phi_5\}$ and $\{\phi_2,\phi_6\}\equiv 
\{\phi_4,\phi_6\}$ correspond to partial blowup by condensation of 
tachyons $T_7$ and $T_3$ respectively. The convex hulls 
$\{\phi_1,\phi_2\}\equiv \{\phi_1,\phi_4\}$ and $\{\phi_1,\phi_3\}$ 
correspond to complete resolutions by condensation of both tachyons 
$T_3$ and $T_7$, one followed by the other, and are related by a flip.

From the phase diagram, we see that the stable phases correspond to 
(i) $\{\phi_2,\phi_6\}\equiv \{\phi_4,\phi_6\}$, condensation of 
$T_3$ alone, and (ii) $\{\phi_1,\phi_2\}\equiv \{\phi_1,\phi_4\}$, 
condensation of $T_3$ followed by a blowup by the now-marginal $T_7$.

A flip transition itself occurs across the $\phi_1$-phase boundary, 
and in the effective subcone $C(0;e_2,e_3,e_4,T_3,T_7)$: it is described 
by the effective D-term equation
\be
|\phi_2|^2+2|\phi_4|^2+3|T_7|^2-|\phi_3|^2-7|T_3|^2={7r_1-3r_2\over 25}=r_f\ .
\ee
The RG flow of this effective FI parameter $r_f$ is $r_f=({-2\over 2\pi}) 
\log {\mu\over\Lambda}$, showing that the flip proceeds in the direction 
approaching the region $7r_1-3r_2>0$, \ie\ the stable phase 
$r_1>{3\over 7} r_2$. For $r_f>0$, this gives a weighted $\BC\BP^2$ 
while for $r_f<0$, we have a (weighted) $\BC\BP^1$. As the phase boundary 
is crossed, the $\BC\BP^1$ blows down and the more stable $w\BC\BP^2$ 
blows up dynamically.

We also see from the auxiliary D-terms that condensation of the tachyon 
$T_7$ in orbifold subcones $\C(0;e_1,e_2,e_3)$ and $C(0;e_1,e_2,e_3,e_5)$ 
occurs across the phase boundaries $\phi_2$ and $\phi_3$ respectively.
The fact that crossing the phase boundary $\phi_2\equiv\phi_4$ 
across the stable phases corresponds to a blowup of the now-marginal 
$T_7$ is reflected in the fact that $r_1-4r_2$ is in fact a marginal 
parameter with no (1-loop) renormalization.

Overall, the stable phase corresponding to blowup by $T_3$ corresponds 
to a $w\BC\BP^3$ expanding out in (RG) time, containing residual 
orbifold singularities on its locus: from the D-term $D_1$, these 
are $C(0;e_2,e_3,e_4,T_3)\equiv \BZ_3 (1,2,2,-1),\ 
C(0;e_1,e_3,e_4,T_3)\equiv \BZ_4 (-1,2,0,-1),\\ 
C(0;e_1,e_2,e_4,T_3) \equiv \BZ_2 (1,0,0,-1),\
C(0;e_1,e_2,e_3,T_3)\equiv \BZ_8 (3,4,2,-1)$, after shifting the 
weights to obtain Type II orbifolds. The now-marginal $T_7$ lies 
in the residual $\BZ_4$ orbifold: its blowup (which is a $w\BC\BP^2$, 
from the D-term $D_{\phi_2}$) gives rise to a further resolution, 
with the resulting space described by the eight coordinate charts 
mentioned earlier. The geometry of these charts and the way they 
interlink with each other in the full space is somewhat richer than
the corresponding structure in $\BC^3/\BZ_N$ orbifolds.

\subsection{Conifold-like $(n_1\ \ n_2\  \ n_3\ -n_4\ -n_5)$ singularities}

Consider toric singularities defined by five lattice points $e_i\in\BZ^4$
satisfying\ $\sum_i Q_i e_i=0$ with\ 
$Q_i=(\bA{ccccc} n_1 & n_2 & n_3 & -n_4 & -n_5 \eA)$,\ 
$n_i>0$ and $\sum_in_i\neq 0$. These are the 4-dim analogs of the 
unstable conifold-like singularities studied in \cite{knconiflips}. 
The maximally supersymmetric subspace in this family 
with $\sum_iQ_i=0$ corresponds to toric Calabi-Yau cones, in some 
sense 4-dim analogs of the 3-dim $L_{abc}$ Calabi-Yau singularities 
\cite{labc}. Some of these (and many other classes of 
singularities) have been discussed in the context of ABJM-like 
theories in \eg\ \cite{zaffaroni,lee,ms1,ms2,kleb2,hanany1,hanany2}.

Let us focus for simplicity on singularities with $n_1=1$: then the 
singularity with\ $Q_i=(\bA{ccccc} 1 & n_2 & n_3 & -n_4 & -n_5 \eA)$ 
can be described as a toric cone defined by the five bounding vectors 
$e_1=(-n_2,-n_3,n_4,n_5), e_2=(1,0,0,0), e_3=(0,1,0,0), e_4=(0,0,1,0), 
e_5=(0,0,0,1)$ in a 4-dim toric lattice. Generically these contain 
lattice points in their interior, which can be interpreted as twisted 
sector tachyons in one of the orbifold subcones.

The D-term equation for these singularities (without any additional 
operators added) is
\be
n_1|\phi_1|^2 + n_2|\phi_2|^2 + n_3|\phi_3|^2 - n_4|\phi_4|^2 
- n_5|\phi_5|^2 = r \ ,
\ee
with the RG flow for the Fayet-Iliopoulos parameter being\ 
$r=({\sum_iQ_i\over 2\pi})\log {\mu\over\Lambda}$. For singularities 
with $n_1+n_2+n_3>n_4+n_5$, \ie\ $\sum_iQ_i>0$, the geometry has an 
intrinsic flow from $r>0$ (in the ultraviolet, $\mu\gg\Lambda$) to 
$r<0$ (in the infrared, $\mu\ll\Lambda$). The $r>0$ phase is a 
weighted $\BC\BP^2$ blown up 
while the $r<0$ phase is a (weighted) $\BC\BP^1$ blown up. Thus the 
dynamical evolution of such a singularity naturally gives rise to 
topology change via the blow-down of a $w\BC\BP^2$ and the blowup of 
a $\BC\BP^1$. For singularities with $n_1+n_2+n_3<n_4+n_5$, \ie\ 
$\sum_iQ_i<0$, the dynamical evolution of the geometry is 
intrinsically from the $r<0$ (blown up $\BC\BP^1$) phase to the 
$r>0$ (blown up $w\BC\BP^2$) phase.

A more detailed sense for the phases can be obtained from the coordinate 
charts describing the $r>0$ and $r<0$ phases: for instance, if $r>0$, 
one of the fields $\phi_1,\phi_2,\phi_3$ must acquire a nonzero vev, 
leaving behind four light fields generically, and similarly for $r<0$.
These give the coordinate charts for the two phases
\bc
$r>0:\qquad (\phi_2,\phi_3,\phi_4,\phi_5), (\phi_1,\phi_3,\phi_4,\phi_5), 
(\phi_1,\phi_2,\phi_4,\phi_5) ,$  \\
$r<0:\qquad (\phi_1,\phi_2,\phi_3,\phi_4), (\phi_1,\phi_2,\phi_3,\phi_5) .$
\ec
The subcones in question are potentially orbifold singularities. For 
instance, with $n_1=1$, using the Smith algorithm of \cite{drmknmrp} or 
otherwise, it is possible to see that\ 
$C(0;e_2,e_3,e_4,e_5)\equiv flat,\ C(0;e_1,e_3,e_4,e_5)\equiv 
\BZ_{n_2}(1,n_3,-n_4,-n_5), C(0;e_1,e_2,e_4,e_5)\equiv 
\BZ_{n_3}(1,n_2,-n_4,-n_5),$\\ $C(0;e_1,e_2,e_3,e_4)\equiv 
\BZ_{n_5}(1,n_2,n_3,-n_4), C(0;e_1,e_2,e_3,e_5)\equiv 
\BZ_{n_4}(1,n_2,n_3,-n_5) $, upto shifts of the orbifold weights by 
the respective orbifold orders, since these cannot be unambiguously 
determined.
It is reasonable then to guess that the Type II GSO projection for 
such a nonsupersymmetric singularity is 
\be\label{GSOn12345}
\sum_iQ_i=n_1+n_2+n_3-n_4-n_5=even ,
\ee
based on the known Type II GSO projection $\sum_ik_i=even$ for 
$\BC^4/\BZ_N (k_1,k_2,k_3,k_4)$ orbifolds, if we make the reasonable 
assumption that the GSO projection is not broken along the RG flow 
describing the decay of the system. Setting $n_1=1$ again for simplicity,
this means\ $n_2+n_3-n_4-n_5=odd$ since $\sum_iQ_i$ is only defined 
$mod\ 2$. For instance, say $n_2=even$: then $n_3-n_4-n_5=odd$, and so
$C(0;e_1,e_3,e_4,e_5)\equiv\BZ_{n_2}(1,n_3,-n_4,-n_5)$ automatically 
admits a Type II GSO projection. Now if say $n_3=odd$, then $n_4+n_5=even$ 
and $C(0;e_1,e_2,e_4,e_5)\equiv\BZ_{n_3}(1,n_2,-n_4,-n_5\pm n_3)$ also 
admits a Type II GSO projection after shifting one of the weights by 
the (odd) order $n_3$. It is straightforward to show that the other 
cases are similarly consistent with the Type II GSO projection.
Finally note also that this is consistent with the supersymmetric 
subclass with $\sum_iQ_i=0$, which do admit Type II descriptions.

In general, there are lattice points in the interior of the cone
$C(0;e_1,e_2,e_3,e_4,e_5)$, representing possible blowup modes of the
singularity. In many cases, with our definitions of the cone bounding 
vectors, such interior lattice points can be thought of as defining
lower order conifold-like singularities: for instance, a lattice point
$e_6\in C(0;e_1,e_3,e_4,e_5)$ defines the subcone
$C(0;e_6,e_2,e_3,e_4,e_5)$ for the lower order conifold-like
singularity with some $Q_i^{(2)}$ satisfying\ $\sum_iQ_i^{(2)}e_i'=0$, 
where $e_i'\in\{e_6,e_2,e_3,e_4,e_5\}$.  This system including the 
interior lattice point can thus be described using a GLSM with an 
enlarged charge matrix $Q_i^a, a=1,2$, where the second row is 
$Q_i^{(2)}$. These lattice points can in general be interpreted as 
twisted tachyons of one or more orbifold subcones above arising in 
the decay of the conifold-like singularity. Thus we expect that not 
all such lattice points will be GSO-preserved, since twisted tachyons 
of the orbifold subcones have nontrivial GSO projections. If an 
interior lattice point, \eg\ $e_6$ above, has to define a Type II 
lower order conifold-like singularity, then we must have\ 
$\sum_iQ_i^2=even$. Thus in general, we obtain the general GSO 
projection 
\be
\sum_i Q_i^a = even, \qquad\qquad\quad a=1,2,\ldots
\ee
for such unstable singularities. This is essentially imposing the 
GSO condition (\ref{GSOn12345}) on each row of the charge matrix 
that represents a conifold-like singularity.

One of the simplest examples of such an unstable singularity is\ 
$Q_i=(1\ 1\ 1\ -1\ -4)$. This decays in the direction of the $\BC\BP^2$ 
blowup, the $\BC\BP^1$ blowup being less stable. The toric cone in fact 
contains no lattice points in its interior so that the final endpoint 
is flat space, the $\BC\BP^2$ being round. The $\BC\BP^1$ blowup 
contains the residual (supersymmetric) terminal singularity 
$\BZ_4 (1,1,1,1)$.

\begin{figure}
\bc
\epsfig{file=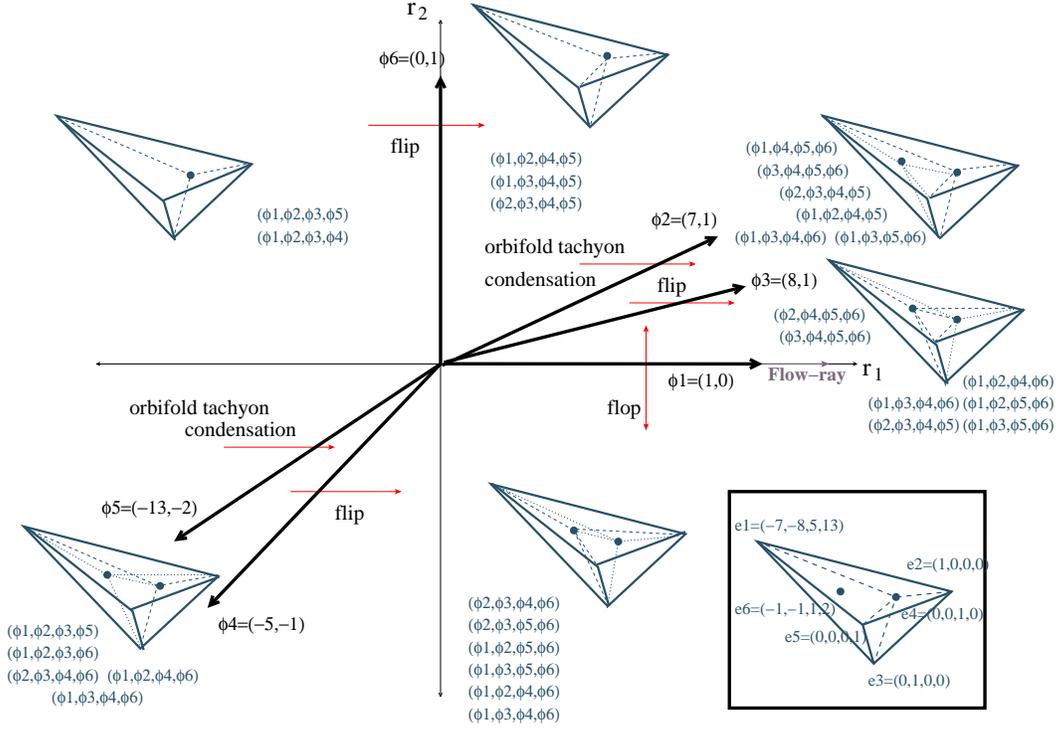, width=14cm}
\caption{{\small The 3-dim hyperplane of the\ $(1\ 7\ 8\ -5\ -13)$ 
flip singularity. The lattice point $e_6$ (and the subdivisions 
thereof) depicted here is really the projection onto this 3-dim 
hyperplane of the actual point (which are in the 4-dim interior of 
the cone).}}
\label{figflip}
\ec
\end{figure}

In general however, the decay structure is more intricate, with 
multiple decay channels due to multiple interior lattice points that 
define lower order orbifold or conifold-like singularities. Thus we 
expect that a high order unstable singularity of this sort will 
typically have a cascade-like decay structure, containing decays to 
lower order singularities amongst its phases. This is indeed the case. 
For instance, consider the singularity\
$Q=(1\ 7\ 8\ -5\ -13)$. The toric cone is defined by the bounding 
vectors\ $e_1=(-7,-8,5,13)$ and $e_2,e_3,e_4,e_5$, as above. Then we see 
that the lattice point\ $e_6=(-1,-1,1,2)={1\over 7} (e_1+e_3+2e_4+e_5)$, 
lies in the interior of the (orbifold) subcone\ $C(0;e_1,e_3,e_4,e_5)
\equiv \BZ_7 (1,8,-5,-13)$. Furthermore using (\ref{Rjlattpt}), 
(\ref{RjPj}), we can recognize this interior lattice point as the 
$j=1$ twisted sector tachyon\ $R_{j=1}\equiv 
({1\over 7},{1\over 7},{2\over 7},{1\over 7})$. 
Note also the relation\ $e_6=(-1,-1,1,2)={1\over 13} (2e_1+e_2+3e_3+3e_4)$,
showing that $e_6$, lying in the interior of $C(0;e_1,e_2,e_3,e_4)\equiv 
\BZ_{13} (1,7,-5,-5)$, represents the $j=2$ sector tachyon\ $R_{j=2}\equiv
({2\over 13},{1\over 13},{3\over 13},{3\over 13})$.\ This suggests that 
this unstable singularity contains a decay to the supersymmetric 
singularity\ $Q=(1\ 1\ 1\ -1\ -2)$ amongst its decay phases (see 
\eg\ \cite{ms1} for a different description of this singularity, 
referred to as $Y^{1,2}(\BC\BP^2)$ there, and other supersymmetric 
singularities): indeed we have the relation\ $e_6+e_2+e_3-e_4-2e_5=0$.
To realize the detailed phase structure of this system (see 
Fig~\ref{figflip}), we use a GLSM with charge matrix 
\be\label{Qia178513}
Q_i^a = \left( \bA{cccccc} 1 & 7 & 8 & -5 & -13 & 0  \\ 0 & 1 & 1 
& -1 & -2 & 1  \\ \eA \right)\ .
\ee

\noindent The phase boundaries are represented by the rays\ 
$\phi_1\equiv (1,0),\ \phi_2\equiv (7,1),\ \phi_3\equiv (8,1),\ 
\phi_4\equiv (-5,-1),\ \phi_5\equiv (-13,-2),\ \phi_6\equiv (0,1)$.\ 
Figure~\ref{figflip} shows the various phases and the corresponding 
subdivisions of the toric cone\footnote{Note that the subcone 
$C(0;e_6,e_2,e_3,e_4,e_5)$ representing the supersymmetric 
$Q=(1\ 1\ 1\ -1\ -2)$ singularity has the following toric subdivisions
for the $w\BC\BP^2$ and $\BC\BP^1$ blowup phases, with\\ 
$r>0:\ (\phi_2,\phi_4,\phi_5,\phi_6), (\phi_3,\phi_4,\phi_5,\phi_6), 
(\phi_2,\phi_3,\phi_4,\phi_5) ,$ \ \ 
$r<0:\ (\phi_2,\phi_3,\phi_5,\phi_6), (\phi_2,\phi_3,\phi_4,\phi_6)$.}.
Although this might be slightly difficult to visualise, it helps to 
note sub-planes defined by three of the six lattice points and the 
relations of the other lattice points relative to the subplanes, 
\eg\ note that $e_4$ and $e_6$ lie on the same side of the 
$\{e_1,e_2,e_3\}$-plane, as can be seen from the relation\ 
$e_6={1\over 13} (2e_1+e_2+3e_3+3e_4)$.\ Ultimately, the phase 
structure is obtained from the D-term equations
\bea
&& -D_1\equiv -D_{\phi_6}
\equiv |\phi_1|^2 + 7|\phi_2|^2 + 8|\phi_3|^2 - 5|\phi_4|^2 
- 13|\phi_5|^2 = r_1\ , \nonumber\\
&& -D_2\equiv -D_{\phi_1}
\equiv |\phi_2|^2 + |\phi_3|^2 + |\phi_6|^2 - |\phi_4|^2 
- 2|\phi_5|^2 = r_2\ , 
\eea
and the auxiliary D-term equations across the other four phase 
boundaries (obtained by eliminating the corresponding field)
\bea
&& -D_{\phi_2}\equiv |\phi_1|^2 + |\phi_3|^2 + 2|\phi_4|^2 + |\phi_5|^2 
- 7|\phi_6|^2 = r_1-7r_2\ , \nonumber\\
&& -D_{\phi_3}\equiv |\phi_1|^2 - |\phi_2|^2 + 3|\phi_4|^2 + 3|\phi_5|^2 
- 8|\phi_6|^2 = r_1-8r_2\ , \nonumber\\
&& -D_{\phi_4}\equiv |\phi_1|^2 + 2|\phi_2|^2 + 3|\phi_3|^2 - 3|\phi_5|^2 
- 5|\phi_6|^2 = r_1-5r_2\ , \\
&& -D_{\phi_5}\equiv 2|\phi_1|^2 + |\phi_2|^2 + 3|\phi_3|^2 + 3|\phi_4|^2 
- 13|\phi_6|^2 = 2r_1-13r_2\ . \nonumber
\eea

Using these, or equivalently the operational method described earlier, 
one finds the various phases with corresponding coordinate charts 
shown in Figure~\ref{figflip}. The 1-loop renormalization of $r_1$ is\ 
$r_1={-2\over 2\pi} \log {\mu\over\Lambda}$ , while $r_2$, with no 
renormalization is a modulus. The system thus flows in the direction 
of the flow-ray\ $F\equiv (1,0)$, which adjoins the stable phases:
these include the decay to the supersymmetric $Q=(1\ 1\ 1\ -1\ -2)$
singularity, the stable phases including the $w\BC\BP^2$ and 
$\BC\BP^1$ blowups thereof. The occasional (rare) decay of the 
system precisely along the flow ray $(1,0)$, \ie\ along 
$r_2=0,\ r_1\ra\infty$, yields the singular point in the moduli 
space of the supersymmetric singularity\footnote{Strictly speaking, 
there is a constant quantum shift in the location of the classical 
singularity at $r_2^{(0)}=0$: this arises from the 1-loop bosonic 
potential (see Appendix C) as 
$t_2^{eff}=t_2^{(0)}+{i\over 2\pi}\sum_iQ_i^2\log|Q_i^2|=0$ 
defining the singular point $r_2^{(0)}=r_2^c$, giving a real 
codimension-2 singularity after including the effects of the 
$\theta$-angle.}.

The ultraviolet of the system is the direction $(-1,0)$ opposite the 
flow-ray, contained in the convex hull $\{ \phi_5,\phi_6\}$: this phase 
corresponds to the unstable $\BC\BP^1$ shrinking in (RG) time. 
Beginning in this phase, the structure of the D-term equations shows 
that the RG evolution to the two stable phases goes through one of 
two possible paths, crossing phase boundaries (i) $\phi_6, \phi_2, 
\phi_3$, or (ii) $\phi_5, \phi_4$. The corresponding phase transitions 
occurring in the process are: (i) a flip occurs across the phase 
boundaries $\phi_6, \phi_3, \phi_4$, (ii) condensation of tachyons 
corresponding to the lattice points $e_6\in C(0;e_1,e_3,e_4,e_5)$ 
and $e_6\in C(0;e_1,e_2,e_3,e_4)$ orbifold subcones occurs across 
$\phi_2$ and $\phi_5$ respectively, while (iii) the phase boundary 
$\phi_1$ corresponds to a flop. The final phases correspond to 
the stable $w\BC\BP^2$ expanding outwards, with possible residual 
singularities on its locus.

\section{$M2$-branes stacked at $\BC^4/\BZ_N$ and 
nonsupersymmetric $AdS_4\times S^7/\BZ_N$ backgrounds}

In recent times, a Chern-Simons field theory description dual 
to the near horizon $AdS_4\times S^7/\BZ_N$ backgrounds obtained from
M2-branes stacked at $\BC^4/\BZ_N (1,1,1,1)$ singularities has been
found in \cite{abjm}. Various generalizations of this to M2-branes
stacked at diverse singularities have been studied in \eg\
\cite{klebanov,terashima,jafferis,imamura,uedayama,lee,ms1,ms2,
kleb2,hanany1,hanany2,hesparks} (see also \eg\ 
\cite{morrisonplesserHorizons,zaffaroni} for some early work on 4-dim
singularities). It would seem along these lines that the dual field
theories to nonsupersymmetric $AdS_4\times S^7/\BZ_N$ backgrounds
obtained from M2-branes stacked at nonsupersymmetric $\BC^4/\BZ_N$
singularities would also be Chern-Simons theories, but
nonsupersymmetric ones. One might imagine that a prescription inspired
by \cite{douglasmoore} for D-branes at orbifold singularities with
image M2-branes on a covering space might also work in this case: this
has been studied for various supersymmetric orbifolds in \eg\
\cite{klebanov,terashima,imamura}. It would be interesting to explore
this further in the nonsupersymmetric context.

We have seen that nonsupersymmetric $\BC^4/\BZ_N$ orbifold
singularities are unstable, with 11-dimensional lifts of closed string
tachyons or moduli ensuring the resolutions of these singularities. We
have used the 11-dimensional reflections of Type IIA worldsheet string
descriptions of these singularities and the phase structure of
associated GLSMs to glean the structure of these singularities. One
might ask if a similar description could be obtained using an M2-brane
probe and Higgsing therein via the couplings of possible
Fayet-Iliopoulos parameters to the M-theory background blowup modes,
along the lines of a D3-brane probe description of lower dimensional
supersymmetric orbifold singularities, as \cite{douglasmoore,dgm}
discussed. Indeed a basis of irreducible loops on the quiver can be
mapped to the gauge invariant monomials obtained (for toric
singularities) in a GLSM description or equivalently the holomorphic
quotient construction (see \eg\ \cite{grantkn} for a more recent
explicit description generalizing \cite{douglasmoore,dgm} to the
context of 3-dim Calabi-Yau $L_{abc}$ singularities, and related work
\cite{hananyLabc,martellisparksLabc}). Investigations of this kind for
M2-branes in the vicinity of various classes of supersymmetric
singularities have been performed in \cite{hanany2}. In the
nonsupersymmetric case, it was found already in \cite{aps} that in
$\BC^2/\BZ_N$ (and unlike $\BC/\BZ_n$), the moduli space of the gauge
theory was not identical to the geometry and its partial
resolutions\footnote{Preliminary investigations (with R. Plesser), in
  the incipient stages of \cite{drmknmrp} seemed to corroborate this
  for $\BC^3/\BZ_N$ singularities.}. It is quite possible that this
will be the case for $\BC^4/\BZ_N$ and M2-brane probes too, once
spacetime supersymmetry is broken. However it would be interesting to
explore this further, especially in the light of the findings of
\cite{klebroib}. Since the Chern-Simons level is related to the
orbifold order for the supersymmetric cases \cite{abjm}, it would seem
that perhaps tachyon condensation (which induces partial resolutions
lowering the order of a singularity) will give rise to flows modifying 
the Chern-Simons level.

Consider now a stack of $k$ $M2$-branes placed at a nonsupersymmetric 
$\BC^4/\BZ_N (k_1,k_2,k_3,k_4)$ singularity, the full M-theory background 
being of the form $\BR^{2,1}\times \BC^4/\BZ_N$. The orbifold can be 
thought of as a cone over $S^7/\BZ_N$. Then for a large number of 
$M2$-branes, taking the near horizon limit gives a nonsupersymmetric 
$AdS_4\times S^7/\BZ_N$ background, the radial direction of the orbifold 
combining with $\BR^{2,1}$ to give $AdS_4$. If the group $\BZ_N$ acts 
freely on the $S^7$, the resulting $S^7/\BZ_N$ space is smooth. This 
implies that there are no fixed points on $S^7$ where localized 
tachyons can arise so that the large flux $AdS_4\times S^7/\BZ_N$ 
limit is apparently tachyon free and thus potentially a stable 
nonsupersymmetric background. In the D-brane context for $AdS_5\times 
S^5/\Gamma$, various aspects were discussed in \cite{kachrusilver,adamseva}.

This however is too quick: there turns out to be a nonperturbative 
gravitational instability \cite{horopolchAdSInstab} of the sort that 
plagues a Kaluza-Klein background manifested by Witten's ``bubble of 
nothing'' \cite{wittenBubble}.
Along the same lines, we expect that the $AdS_4\times S^7/\BZ_N$ can 
be recast as a Kaluza-Klein compactification over an $S^1$ of 
$AdS_4\times w\BC\BP^3$, for some appropriate weighted projective 
3-space $w\BC\BP^3$ and $S^1$ periodicity $\sim {1\over N}$. 
This then will give rise to a similar bubble-of-nothing 
instability for nonsupersymmetric $AdS_4\times S^7/\BZ_N$ backgrounds, 
which are then expected to decay rapidly: as pointed out in the 
$AdS_5\times S^5/\BZ_N$ case \cite{horopolchAdSInstab}, the decay rate 
for a conformal theory (no scale) must be zero or infinite, and 
the integral over the radial coordinate diverges.

An interesting detail here requires understanding the fermion boundary
conditions across the $S^1$ for these nonsupersymmetric cases. It
would seem that the structure here for general orbifold weights is
intricate from the supergravity point of view, since as we run through
possible weights, for precisely the supersymmetric values (which are
stable), the instanton must not exist. For instance, $\BZ_N (1,3,5,7)$
is a supersymmetric singularity, while $\BZ_N (1,5,7,9)$ is not, 
and $\BZ_N (1,5,7,11)$ again is supersymmetric. From this point of 
view, strictly speaking it is not obvious if every nonsupersymmetric 
$AdS_4\times S^7/\BZ_N$ background necessarily admits a KK-instanton
(manifest in supergravity) that mediates its decay.

The noncompact case apart, as for $AdS_5\times S^5/\BZ_N$ interestingly 
pointed out by \cite{kachrutrivediAdSorbs} (and already noted in
\cite{horopolchAdSInstab}), the decay rate in the present case is
strictly infinite only if the throat is infinitely long corresponding
to M2-branes stacked at a noncompact $\BC^4/\BZ_N$ singularity. If the
throat is instead embedded in an orbifold of an appropriate compact
Calabi-Yau 4-fold, then the divergence of the decay rate is regulated
by the ultraviolet region, which is the compact Calabi-Yau
orbifold. With the instanton action being $B$, cutting off the field 
theory at a scale $\Lambda$, \ie\ at a radial coordinate 
$r_{UV}\sim R^2\Lambda$, $R$ being the AdS radius, gives the total 
decay rate per unit $2+1$-dim field theory volume as
\be
\Gamma \sim\ e^{-B} \int^{r_c} dr r^2 \sim\ e^{-B} \Lambda^3\ ,
\qquad\quad B\sim {r_0^9\over G_{11}} \sim\ {k^{3/2}\over N^9}\ ,
\ee
where $r_0\sim {R\over N}$ is the radial coordinate value where the 
instanton is capped off. Thus for fixed orbifold order $N$ and large 
number $k$ of M2-branes, the instanton action is large, giving a 
small decay rate, as for $AdS_5\times S^5/\BZ_N$ argued in 
\cite{kachrutrivediAdSorbs}.
In this context of a throat cutoff at some $r_{UV}$ with a compact 
space, perhaps such backgrounds provide useful stable 
nonsupersymmetric $AdS_4\times S^7/\BZ_N$ throats in M-theory.

\vspace{3.5mm}
{\small
{\bf Acknowledgements:} I have benefitted from conversations with 
S. Govindarajan, N. Suryanarayana and S. Trivedi. I thank the 
hospitality of the CERN Theory Group over the ``String Phenomenology'' 
Workshop, Aug 2008, during the incipient stages of this work, and 
S. Mukhi and H. Singh for discussions at the time. This work is 
partially supported by a Ramanujan Fellowship, DST, Govt. of India.}

\vspace{7mm}

\appendix
\section{Aspects of the $\BC^4/\BZ_N$ spectrum}

The twisted sector spectrum of the $\BC^4/\BZ_N (k_1,k_2,k_3,k_4)$ 
orbifold conformal field theory, classified using the representations 
of the $(2,2)$ superconformal algebra, has a product-like structure. 
The worldsheet supercurrents for each complex plane are 
$G^+_i=\psi_i^*\del X_i$ or $G^-_i=\psi_i\del X_i^*$, with the $U(1)$ 
currents being $J_i=\psi_i^*\psi_i$. Consider a twist sector $j$, with 
boundary conditions 
$X^i(\sigma+2\pi,\tau)=e^{2\pi i jk_i/N} X^i(\sigma,\tau)$. 
The worldsheet fermions have half-integral moding, 
$\psi_i(z)=\sum_{r\in\BZ+{1\over 2}} 
\psi_{r+\{{jk_i\over N}\}}/z^{r+\{{jk_i\over N}\}+{1\over 2}}$, and
$\psi^*_i(z)=\sum_{r\in\BZ+{1\over 2}} 
\psi^*_{r-\{{jk_i\over N}\}}/z^{r+\{{jk_i\over N}\}-{1\over 2}}$: thus\ 
$\psi_{\{{jk_i\over N}\}-{1\over 2}}$ (or $\psi^*_{{1\over 2}-\{{jk_i\over N}\}}$) 
changes from a creation (or annihilation) operator to an annihilation 
(or creation) operator as $\{{jk_i\over N}\}$ grows greater than 
${1\over 2}$, with respect to the twist ground state $|0\rangle_j$, 
annihilated by all operators with positive moding (see \eg\ 
\cite{sinGSO} for a lucid discussion on this). Thus for a complex 
plane-$i$, $|0\rangle_j$ changes from being a chiral state with 
$G^+_{-{1\over 2}}|0\rangle_j=0$\ to an anti-chiral state with 
$G^-_{-{1\over 2}}|0\rangle_j=0$\ (note that\ 
$G^+_{-{1\over 2}}=\psi^*_{{1\over 2}-\{{jk_i\over N}\}} \al_{\{{jk_i\over N}\}-1} 
+ \ldots$ and $G^-_{-{1\over 2}}=\psi_{\{{jk_i\over N}\}-{1\over 2}} 
\al^*_{-\{{jk_i\over N}\}} + \ldots$, the $\al$'s being the operators 
entering in the worldsheet boson mode expansions). Likewise for 
$0<\{{jk_i\over N}\}<{1\over 2}$ , the first excited state is
$\psi_{\{{jk_i\over N}\}-{1\over 2}} |0\rangle_j$ and is antichiral, while 
for $\{{jk_i\over N}\}>{1\over 2}$ , the first excited state is 
$\psi^*_{{1\over 2}-\{{jk_i\over N}\}} |0\rangle_j$ is chiral.
The product structure of the orbifold conformal field theory implies 
that the spectrum of ground and first excited states can be segregated 
into various chiral and antichiral rings comprising states that are 
chiral under either $G^+_i$ or $G^-_i$ for each complex plane: thus 
\eg\ the $(c_{X_1},c_{X_2},c_{X_3},c_{X_4})$ ring consists of states 
chiral under $\sum_{i=1}^4 G^+_i$, while \eg\ the 
$(c_{X_1},c_{X_2},c_{X_3},a_{X_4})$ ring consists of states chiral 
under $\sum_{i=1}^3 G^+_i + G^-_4$. This gives sixteen chiral and 
anti-chiral rings in eight conjugate pairs.

The zero point energy for the left-moving modes for a single orbifolded 
complex plane is (with $a_i=\{{jk_i\over N}\}$)
\bea
E_0^i &=& {1\over 2}\sum_{n=0}^\infty (n+a_i) + 
{1\over 2}\sum_{n=0}^\infty (n-a_i) 
- {1\over 2}\sum_{n=0}^\infty (n+{1\over 2}+a_i) 
- {1\over 2}\sum_{n=0}^\infty (n+{1\over 2}-a_i) \nonumber\\
&=& {1\over 2} a_i - {1\over 8}\ , \qquad\qquad\qquad 0<a_i<{1\over 2}\ ,
\eea
which, adding up, gives\ 
\be
E_0={1\over 2} \sum_i \{{jk_i\over N}\} - {1\over 2}\ , \qquad 
0<\{{jk_i\over N}\}<{1\over 2}\ ,
\ee
where we have used 
the regularized sum\ 
$\sum_{n=0}^\infty (n+a)={1\over 24}-{1\over 8} (1-2a)^2$ .
If say ${1\over 2}<\{{jk_4\over N}\} <1$, then with the new 
creation-annihilation operators entering, the zero point energy is 
modified to
\be
E_0' = {1\over 2} \sum_{i\neq 4} \{{jk_i\over N}\} - {1\over 2} 
- {1\over 2} \left[ -\left({1\over 2} - \{{jk_4\over N}\} \right)
+ \left(\{{jk_4\over N}\} - {1\over 2} \right) \right] 
= {1\over 2} \sum_{i\neq 4} \{{jk_i\over N}\} - {1\over 2} 
- {1\over 2} \{{jk_4\over N}\} \ ,
\ee
which can be recast as $E_0'={1\over 2} \sum_i \{{jk_i'\over N}\}-{1\over 2}$, 
for $k_i'=(k_1,k_2,k_3,-k_4)$. Thus the conformal weights and R-charges 
of the twist ground states satisfy
\be
E_0 = \Delta - {1\over 2}\ , \qquad \Delta=\pm {1\over 2} R\ .
\ee

We now describe the RNS partition functions: these are generalizations of 
those for nonsupersymmetric $\BC^3/\BZ_N$ singularities \cite{drmknmrp}.
The Type 0 string on $\BC^4/\BZ_N (k_1,k_2,k_3,k_4)$ has a diagonal 
GSO projection that ties together the left and right movers: it has 
the partition function 
\bea
&& Z = {1\over 2N} \sum_{j,l=0}^{N-1}\ 
\prod_{i=1}^4\ \left| {\eta(\tau)\over \theta 
[^{1/2+jk_i/N}_{1/2+lk_i/N}](0,\tau)} \right|^2 
\Biggl[  \biggl| \prod_{i=1}^4 \theta [^{jk_i/N}_{lk_i/N}]\biggr|^2 + 
\biggl| \prod_{i=1}^4 \theta [^{jk_i/N}_{lk_i/N+1/2}]\biggr|^2 \nonumber\\
&& \qquad\qquad\qquad\qquad\qquad\qquad 
+ \biggl| \prod_{i=1}^4 \theta [^{jk_i/N+1/2}_{lk_i/N}]\biggr|^2 
\pm \biggl| \prod_{i=1}^4 \theta [^{jk_i/N+1/2}_{lk_i/N+1/2}]\biggr|^2 
\Biggr] ,
\eea
which exists for any $k_i,N$. 
On the other hand, the 1-loop partition function on a 
$\BC^4/\BZ_N\ (k_1,k_2,k_3,k_4)$ orbifold for a Type II string with 
separate GSO projections on the left and right movers is given by the 
sum over twisted sectors as 
\be\label{IIpf}
Z = {1\over 4N} \sum_{j,l=0}^{N-1}\ 
\prod_{i=1}^4\ \biggl| {\eta(\tau)\over 
\theta [^{1/2+jk_i/N}_{1/2+lk_i/N}](0,\tau)}\biggr|^2 
\biggl| {\zeta^j_l \over |\eta^4(\tau)} \biggr|^2\ ,
\ee
\be
\zeta^j_l = \prod_{i=1}^4 \theta [^{jk_i/N}_{lk_i/N}] 
- e^{-i\pi\sum_i {jk_i\over N}} 
\prod_{i=1}^4 \theta [^{jk_i/N}_{lk_i/N+1/2}] 
- \prod_{i=1}^4 \theta [^{jk_i/N+1/2}_{lk_i/N}] 
- e^{-i\pi\sum_i {jk_i\over N} } \prod_{i=1}^4 
\theta [^{jk_i/N+1/2}_{lk_i/N+1/2}]\ .
\ee
$\zeta_j^l$ contains the sum over spin structures for the j-th twisted 
sector twisted by $g^l$ in the ``time'' direction. The terms in Z are 
recognized as the contributions from the twisted bosons in the 
orbifolded complex dimensions and the fermionic contributions. 

This is modular invariant (in particular under the S-transformation) 
if the phase from the third term above satisfies\ \
$e^{i\pi\sum_i{lk_i\over N}}=
e^{-i\pi\sum_i{(N-l)k_i\over N}}$,\ in other words, 
\be
\sum_i {(N-l)k_i\over N} = -\sum_i {lk_i\over N} + {\rm even}\qquad
\Longrightarrow\qquad \sum_i k_i=even .
\ee
We can now expand the Type II partition function (\ref{IIpf}) by 
expanding the $\theta$-functions as\ 
$\theta [^a_b] (0,\tau)=\sum_{n=-\infty}^\infty q^{{1\over 2} (n+a)^2} 
e^{2\pi i (n+a) b},\ q=e^{2\pi i\tau}$, to realize the GSO projection on 
the twisted states: we obtain the projector 
\be\label{projcccc}
1 - (-1)^{\sum_i [jk_i/N]}
\ee
for the ground states in the sector where $\{ {jk_i\over N} \} < 
{1\over 2}$, \ie\ the $(c_{X_1},c_{X_2},c_{X_3},c_{X_4})$ ring. This is 
a projector onto twisted states with \ $\sum_i [jk_i/N]=E_j=odd$. In 
the untwisted $j=0$ sector, this is in accord with the usual $[1+(-1)^F]$ 
GSO projection that removes the (bulk) closed string tachyon (after 
accounting for a $(-1)$ from the ghost contribution to the worldsheet 
$(-1)^F$). 
Consider now \eg\ the sector where $\{{jk_4\over N}\}>{1\over 2}$ 
with other $\{{jk_i\over N}\}<{1\over 2}$. Then we obtain the projector 
\be\label{projccca}
1 - (-1)^{(\sum_i [jk_i/N] - 1)}
\ee
for the ground states (which are in the $(c_{X_1},c_{X_2},c_{X_3},a_{X_4})$ 
ring), \ie\ $E_j=\sum_i [jk_i/N]=even$. The chiral operators $X_j$ 
are obtained as the excited state with one extra fermion number from 
$\psi_4$ which therefore have the GSO projection 
$\sum_i [jk_i/N]=E_j^{ccca}=odd$, as before. Likewise if two of 
$\{{jk_i\over N}\}>{1\over 2}$, we have $E_j=odd$ for the ground 
states so that the $X_j$, obtained with one extra fermion number in the 
two sectors, again have $E_j=odd$ and so on. Thus the GSO exponent for 
the chiral operators $X_j$ is $E_j=\sum_i [jk_i/N]=odd$. 

Thus we see that the GSO exponents for the various rings are\
\bea\label{EjGSO}
E_j=odd, && (c_{X_1},c_{X_2},c_{X_3},c_{X_4}), 
(c_{X_1},c_{X_2},a_{X_3},a_{X_4}), (c_{X_1},a_{X_2},c_{X_3},a_{X_4}), 
(c_{X_1},a_{X_2},a_{X_3},c_{X_4}) , \nonumber\\
&& (a_{X_1},c_{X_2},c_{X_3},a_{X_4}), (a_{X_1},c_{X_2},a_{X_3},c_{X_4}), 
(a_{X_1},a_{X_2},c_{X_3},c_{X_4}), (a_{X_1},a_{X_2},a_{X_3},a_{X_4}) ,
\nonumber\\
E_j=even, && (c_{X_1},c_{X_2},c_{X_3},a_{X_4}), 
(c_{X_1},c_{X_2},a_{X_3},c_{X_4}), (c_{X_1},a_{X_2},c_{X_3},c_{X_4}), 
(c_{X_1},a_{X_2},a_{X_3},a_{X_4}) ,\nonumber\\
&& (a_{X_1},c_{X_2},c_{X_3},c_{X_4}), (a_{X_1},c_{X_2},a_{X_3},a_{X_4}), 
(a_{X_1},a_{X_2},a_{X_3},c_{X_4}), (a_{X_1},a_{X_2},c_{X_3},a_{X_4}) .
\nonumber\\
\eea
Note that this is consistent with both a twist field and its conjugate 
field (in the conjugate ring) being GSO-preserved. For instance a 
$(c_{X_1},c_{X_2},c_{X_3},c_{X_4})$-ring twist field operator 
$X_j=\prod_{i=1}^4 X^i_{\{jk_i/N\}}=\prod_{i=1}^3 \sigma_{\{jk_i/N\}} 
e^{i\{jk_i/N\} (H_i-{\bar H}_i)}$\ (in the $(-1,-1)$ picture) 
has its conjugate field 
$X_j^*=\prod_{i=1}^4 (X^i_{\{jk_i/N\}})^*=\prod_{i=1}^4 (X^i_{\{-(N-j)k_i/N\}})^*
=\prod_{i=1}^4 (X^i_{1-\{(N-j)k_i/N\}})^*$,
lying in the $(N-j)$-th twist sector in the conjugate ring 
$(a_{X_1},a_{X_2},a_{X_3},a_{X_4})$. So we see that if $X_j$ is preserved, 
\ie\ $E_j=odd$, then $E_{N-j}=\sum_i [{(N-j)k_i\over N}]=-E_j-4+even=odd$ 
too, preserving the conjugate field in the conjugate ring too.

We can equivalently understand this by ``engineering'' a chiral Type II 
GSO projection for a $\BC^4/\BZ_N (k_1,k_2,k_3,k_4)$ orbifold, consistent 
with that for lower dimensional $\BC^3/\BZ_N$ orbifolds \cite{drmknmrp}.
Complexify the eight transverse untwisted fermions as\ $\psi_i=e^{iH_i},
\ i=0,1,2,3$, and consider a symmetry acting on the untwisted (complex) 
fermions and the twist fields via $H_i \ra H_i + a_i \pi$, 
\be\label{gso1}
\psi_i \ra \psi_i\ {\rm e}^{ia_i\pi}\equiv \psi_i (-1)^{a_i}, \qquad\quad
X_j \ra X_j\ {\rm exp} \Bigl[i\pi \sum_i a_i \Bigl\{ {jk_i\over N} 
\Bigr\} \Bigr]  \equiv X_j\ (-1)^{E_j}\ .
\ee
This defines a $(-1)^{F_L}$ $\BZ_2$ action on the untwisted sector thus 
eliminating the bulk tachyon only if the $a_i$ are odd integers. The 
action on the twisted states $X_j$ is a well-defined $\BZ_2$ if the 
exponent $E_j$ is an integer. This GSO exponent can be written as
$E_j=\sum_ia_i\{{jk_i\over N}\}={j\over N}\sum_ia_ik_i-\sum_ia_i[{jk_i\over N}]$.
Thus $E_j$ is integral if we have $a_i=odd$ satisfying\ $\sum_ia_ik_i=0\ 
(mod\ 2N)$.\ If $\sum_ik_i=odd$, we see that 
$\sum_{i=1}^4a_ik_i=\sum_{i=1}^3(a_i-a_4)k_i+odd=odd$,\ since the first 
three terms (containing differences of odd integers) are even. Thus 
no odd $a_i$ exist satisfying\ $\sum_ia_ik_i=0 (mod\ 2N)$ if $\sum_ik_i=odd$\ 
(assuming even $N$: for odd $N$, one can shift one of the weights to 
make $\sum_ik_i=even$).\\
For $\sum_ik_i=even$, we thus have\ 
$E_j=\sum_ia_i\{{jk_i\over N}\}=\sum_ia_i[{jk_i\over N}]$, and\ 
$\sum_ia_ik_i=0 (mod\ 2N)$. 
Thus for a twisted state $T$ with R-charge $R=(r_1,r_2,r_3,r_4)$ in 
the orbifold $\BC^4/\BZ_N\ (k_1,k_2,k_3,k_4)$, the GSO exponent is\ 
$E=\sum_ia_ir_i$\ with\ $a_i=odd$ and $\sum_ia_ik_i=0 (mod\ 2N)$.\\
Now for a Type II orbifold $\BC^4/\BZ_N (1,p,q,r)$, we have\ 
$p+q+r=odd$:\ then\ $a_1=p+q+r, a_2=a_3=a_4=-1$, satisfy\ 
$\sum_ia_ik_i=a_1+a_2p+a_3q+a_4r=0 (mod\ 2N)$. This gives the GSO 
exponent\ 
$E_j=[{jp\over N}]+[{jq\over N}]+[{jr\over N}]=\sum_i[{jk_i\over N}]$.

\section{The Maple program}

The Maple code we have used is sufficiently simple and we give it here:
{\footnotesize
\begin{verbatim}
w := [1,7,9,11];
k := [[w[1],w[2],w[3],w[4]], [w[1],w[2],w[3],-w[4]], [w[1],w[2],-w[3],w[4]], 
[w[1],-w[2],w[3],w[4]], [w[1],w[2],-w[3],-w[4]], [w[1],-w[2],-w[3],w[4]], 
[w[1],-w[2],w[3],-w[4]], [w[1],-w[2],-w[3],-w[4]]];
for N from 2 to 400 do
 for j from 1 to 8 do
  for l from 1 to N-1 do
  for i from 1 to 4 do 
   rl[j,i] := l*k[j,i]/N - floor(l*k[j,i]/N):
   fl[j,i] := floor(l*k[j,i]/N):
  end do:
   Rl := sum('rl[j,i]', 'i=1..4'):
   El := sum('fl[j,i]', 'i=1..4'):
 if (Rl < 1) then if (type(El,odd))  then 
  print(N,j,l,[rl[j,1],rl[j,2],rl[j,3],rl[j,4]],Rl,El,'tachyon')
  end if: end if: 
 if (Rl = 1) then if (type(El,odd))  then 
  print(N,j,l,[rl[j,1],rl[j,2],rl[j,3],rl[j,4]],Rl,El,'marginal')
  end if: end if: 
end do:
end do:
end do;
\end{verbatim} }
This is a significantly improved and simplified version of a code written 
for $\BC^3/\BZ_N$ towards the completion of \cite{drmknmrp}. Once we 
input the weights $w=(w_1,w_2,w_3,w_4)$, the program calculates the 
various GSO-preserved twisted sector R-charges in the eight pairs of 
chiral and antichiral rings of a $\BC^4/\BZ_N (w_1,w_2,w_3,w_4)$ orbifold 
for $N\leq 400$, and lists tachyons and moduli that arise as $N$ 
increases. Thus if a particular orbifold order $N_0$ does not appear in 
the program output, there are no tachyons or moduli in its spectrum, \ie\
it is terminal. Various modifications of this can be easily written to
accomodate a different range for $N, k_i$, specific rings, or calculate 
\eg\ the spectrum of a Type 0 orbifold with a diagonal GSO projection.

\section{Some aspects of GLSMs}

This subsection is essentially a direct generalization (primarily for 
completeness) of the techniques described in \cite{drmkn,knconiflips} 
to the 4-dim singularities in question here. 
The full phase structure of a (noncompact) $\BC^4/\BZ_N$ orbifold 
geometry (such as those discussed in this paper) with $n$ tachyons 
is obtained by studying the Higgs branch of the moduli space of an 
enlarged gauged linear sigma model (GLSM), admitting $(2,2)$ worldsheet 
supersymmetry, with gauge group $U(1)^n$, $4+n$ chiral superfields 
$\Psi_i$ and $n$ Fayet-Iliopoulos parameters $r_a$. The action of such a 
GLSM (in conventions of \cite{wittenphases,morrisonplesserInstantons}) is
\be
S = \int d^2 z\ \biggl[ d^4 \theta\ \biggl( {\bar \Psi_i} e^{2Q_i^a 
V_a} \Psi_i - {1\over 4e_a^2} {\bar \Sigma_a} \Sigma_a \biggr) + 
\Rea\biggl( i t_a\int d^2 {\tilde \theta}\ \Sigma_a  \biggr) \biggr]\ ,
\ee
where summation on the index $a=1,\ldots, n$ is implied. The\ $t_a = 
ir_a + {\theta_a\over 2\pi}$ \ are Fayet-Iliopoulos parameters and 
$\theta$-angles for each of the $n$ gauge fields ($e_a$ being the 
gauge couplings). The twisted chiral superfields $\Sigma_a$ (whose 
bosonic components are complex scalars $\sigma_a$) represent 
field-strengths for the gauge fields. The action of the $U(1)^n$ 
gauge group on the $\Psi_i$ is given in terms of the $n\times 
(4+n)$ charge matrix $Q_i^a$ above as 
$\Psi_i \ra e^{i Q_i^a\lambda} \Psi_i\ , 
a=1,\ldots,n .$
For the conifold-like singularities, we have $5+n$ superfields and 
$n+1$ FI parameters, with a gauge group $U(1)^{n+1}$ and a
$(n+1)\times (5+n)$ charge matrix: the $n$ superfields in this case 
represent lattice points in the interior of one of the subcones in 
the original singular cone $C(0;e_1,e_2,e_3,e_4,e_5)$. For instance, 
we have $Q_i^a$ in (\ref{Qia2517911}) for the orbifold with $n=2$ 
tachyons, and $n=1$ interior lattice point in (\ref{Qia178513}) for 
the conifold-like singularity. Such a charge matrix only specifies 
the $U(1)^n$ action up to a finite group, due to the possibility of 
a $\BQ$-linear combination of the rows of the matrix also having 
integral charges. The specific form of $Q_i^a$ is chosen to 
conveniently illustrate specific geometric substructures, \eg\ the 
tachyons contained within the orbifold, or subcones representing 
lower order conifold-like singularities.
The variations of the $n$ independent FI parameters control the 
vacuum structure of the theory. The space of classical ground states 
of this theory can be found from the bosonic potential\
$U = \sum_a {(D_a)^2\over 2e_a^2} + 2\sum_{a,b} {\bar \sigma}_a 
\sigma_b \sum_i Q_i^a Q_i^b |\Psi_i|^2 .$
Then $U=0$ requires $D_a=0$: solving these for $r_a\neq 0$ gives 
expectation values for the $\Psi_i$, which Higgs the gauge group down 
to some discrete subgroup and lead to mass terms for the $\sigma_a$ 
whose expectation values thus vanish. The classical vacua of the 
theory are then given in terms of solutions to the D-term equations
\be
{-D_a\over e^2} = \sum_i Q_i^a |\Psi_i|^2 - r_a = 0\ , 
\qquad a=1,\ldots,n\ .
\ee
At the generic point in $r$-space, the $U(1)^n$ gauge group is 
completely Higgsed, giving collections of coordinate charts that 
characterize in general distinct toric varieties. In other words, 
this $n$-parameter system admits several ``phases'' (convex hulls 
in $r$-space, defining the secondary fan) depending on the values 
of the $r_a$. At boundaries between these phases where some (but not 
all) of the $r_a$ vanish, some of the $U(1)$s survive giving rise to
singularities classically.  Each phase is an endpoint since if left
unperturbed, the geometry can remain in the corresponding resolution 
indefinitely (within this noncompact approximation): in this sense, 
each phase is a fixed point of the GLSM RG flow. However some
of these phases are unstable while others are stable, in the sense
that fluctuations (\eg\ blowups/flips of cycles stemming from
instabilities) will cause the system to run away from the unstable 
phases towards the stable ones. This can be gleaned from the 1-loop 
renormalization of the FI parameters
\be\label{flow}
r_a = \bigg({\sum_iQ_i^a\over 2\pi}\bigg) \log {\mu\over \Lambda}\ ,
\ee
where $\mu$ is the RG scale and $\Lambda$ is a cutoff scale where 
the $r_a$ are defined to vanish. Energy scales here are defined 
relative to that set by the gauge coupling $e$, which has mass 
dimension one in 2-dim here. The full GLSM RG flow first goes from 
free gauge theory in the ultraviolet $\mu\gg e$ through $\mu\ll e$ but 
with nontrivial dynamics w.r.t. $\Lambda$. Thus in the low energy 
regime $\mu\ll e$, fluctuations transverse to the moduli space cost 
energy and the low-lying fluctuations are simply scalars defining the 
moduli space of the theory: thus the GLSM RG flows approximate the 
nonlinear ones and a geometric description emerges, given by a 
nonlinear sigma model on the moduli space. With attention restricted 
to quasi-topological observables in an appropriate topologically 
twisted A-model, the gauge coupling $e^2$ itself is not crucial to 
the discussion.

A generic linear combination of the gauge fields coupling to a 
linear combination $\sum_a\al_ar_a$ of the FI parameters, the 
$\al_a$ being arbitrary real numbers, has a 1-loop running whose 
coefficient vanishes if
\be\label{ra1loop}
\sum_{a=1}^{n} \sum_{i=1}^{n+4} \al_a Q_i^a = 0\ ,
\ee
in which case the linear combination is marginal. 
This equation defines a codimension-one hyperplane perpendicular to 
a ray, called the Flow-ray, emanating from the origin and passing 
through the point $(-\sum_i Q_i^1, -\sum_i Q_i^2, \ldots, -\sum_i 
Q_i^{n})$ in $r$-space which has real dimension $n$ (for orbifolds; 
for conifold-like singularities, $a=1,\ldots,n+1, i=1,\ldots,5+n$). 
Using the redefinition\ 
${Q_i^a}'\equiv(\sum_iQ_i^1)Q_i^a-(\sum_iQ_i^a)Q_i^1 , \ a\neq 1$, 
we see that\ $\sum_i{Q_i^a}'=(\sum_iQ_i^1)(\sum_iQ_i^a)-(\sum_iQ_i^a)
(\sum_iQ_i^1)=0$, \ for $a\neq 1$, \ so that the FI parameters coupling 
to these redefined $n-1$ gauge fields have vanishing 1-loop running. 
Thus there is a single relevant direction (along the flow-ray) and 
an $(n-1)$-dimensional hyperplane of the $n-1$ marginal directions in 
$r$-space. By studying various linear combinations $\sum_a\al_ar_a$, 
we see that the 1-loop RG flows drive the system along the single 
relevant direction to the (stable) phases in the large $r$ regions of 
$r$-space, \ie, $r_a\gg 0$ (if none of the $r_a$ is marginal), 
that are adjacent to the Flow-ray 
$F\equiv(-\sum_iQ_i^1,-\sum_iQ_i^2,\ldots,-\sum_iQ_i^{n})$, 
or contain it in their interior. 

The direction precisely opposite to the Flow-ray, \ie\
$-F\equiv(\sum_iQ_i^1,\sum_iQ_i^2,\ldots,\sum_iQ_i^{n})$, defines
the ultraviolet of the theory. This ray $-F$ will again lie either in 
the interior of some one convex hull or adjoin multiple convex hulls.
It corresponds to the maximally unstable direction which is the 
unresolved orbifold phase for orbifold geometries or generically the 
unstable $w\BC\BP^2$ or $w\BC\BP^1$ resolution for the conifold-like 
singularities.

We restrict attention to the large $r_a$ regions in the phase diagrams
(in Figures~\ref{figorb}, \ref{figflip}), thus ignoring worldsheet
instanton corrections: this is sufficient for understanding the phase
structure, and consistent for initial values of $r_a$ whose components
in the marginal directions lie far from the center of the marginal
$(n-1)$-plane.

The 1-loop renormalization of the FI parameters can be expressed
\cite{wittenphases, wittenIAS, morrisonplesserInstantons} in terms of
a perturbatively quantum-corrected twisted chiral superpotential for
the $\Sigma_a$ for a general $n$- or $n+1$-parameter system, obtained by
considering the large-$\sigma$ region in field space and integrating
out those scalars $\Psi_i$ that are massive here (and their
expectation values vanish energetically). This leads to the modified
potential\
$U(\sigma) = {e^2\over 2} \sum_{a=1}^{n} \bigg| i{\hat t}_a - 
{\sum_{i=1}^{4+n} Q_i^a \over 2\pi} (\log (\sqrt{2} \sum_{b=1}^{n} 
Q_i^b \sigma_b/\Lambda) + 1) \bigg|^2 $
(for orbifolds). The singularities predicted classically at the 
locations of the phase boundaries arise from the existence of low-energy 
states at large $\sigma$. The physics for the nonsupersymmetric cases 
here is somewhat different from the cases where $\sum_iQ_i^a=0$ for 
all $a$, as discussed in general in \cite{wittenphases, wittenIAS, 
morrisonplesserInstantons} (and for 3-dim singularities in 
\cite{drmkn,knconiflips}). Consider the vicinity of such a singularity 
at a phase boundary but 
far from the (fully) singular region where all $r_a$ are zero, and 
focus on the single $U(1)$ (with say charges $Q_i^1$) that is 
unbroken there (\ie\ we integrate out the other $\sigma_a,\ a\neq 1$, 
by setting them to zero). Now if $\sum_iQ_i^1=0$ (\ie\ unbroken
spacetime supersymmetry), then there is a genuine singularity when
$U(\sigma)={e^2\over 2}|i{\hat t}_1-{1\over 2\pi}
\sum_iQ_i^1\log|Q_i^1||^2=0$, and if $\sum_iQ_i^a=0$ for all $a$, 
this argument can be applied to all of the $U(1)$s. However for the 
nonsupersymmetric cases here, we have $\sum_iQ_i^a\neq 0$: so if say 
$\sum_iQ_i^1\neq 0$ (with the other $Q_i^a$ redefined to ${Q_i^a}'$ 
with $\sum_i{Q_i^a}'=0$), then along the single relevant direction
where $\sum_iQ_i^1\neq 0$, the potential energy has a $|\log
\sigma_1|^2$ growth. Thus the field space accessible to very low-lying
states is effectively compact (for finite worldsheet volume) and there
is no singularity for any $r_a,\theta_a$, along the RG flow: in other
words, the RG flow is smooth along the relevant direction for all
values of $t_1$, and the phase boundaries do not indicate
singularities. In these nonsupersymmetric cases, Coulomb branch 
vacua arise in the IR of the flow 
\cite{wittenphases,wittenIAS,GLSMcoulombbranch}.

\section{Phases of supersymmetric singularities}

Here we apply the techniques we have used here to glean the phase
structure of supersymmetric orbifold and conifold-like singularities.
We will elucidate two examples.

The orbifold $\BC^4/\BZ_{11} (1,1,5,7)$ has moduli arising in the 
$ccca$-ring, with spectrum equivalent to $\BZ_{11} (1,1,5,-7)$: these
are the GSO-preserved twisted sector states with R-charges\ $R_1\equiv 
({1\over 11},{1\over 11},{5\over 11},{4\over 11})=1$ and $R_3\equiv 
({3\over 11},{3\over 11},{4\over 11},{1\over 11})=1$. It has the toric 
cone defined by $e_1=(11,-1,-5,7), e_2=(0,1,0,0), e_3=(0,0,1,0), 
e_4=(0,0,0,1)$. Using (\ref{Rjlattpt}) and (\ref{RjPj}), the moduli 
can be seen to correspond to the lattice points $P_1=(1,0,0,1), 
P_3=(3,0,-1,2)$, lying on the marginality hyperplane. We have also 
the relations\ $P_1={1\over 3} (P_3+e_3+e_4)$ and 
$P_3={1\over 4} (e_1+e_2+P_1+e_3)$, which indicate $P_1$ lies on 
the $\{P_3,e_3,e_4\}$ plane, and $P_3\in C(0;P_1,e_1,e_2,e_3)$. From 
the point of view of the toric cone, the various phases arise from 
the different ways of blowing up the singularity by these moduli. 
These phases can be described by a GLSM with charge matrix
\be\label{Qia111157}
Q_i^a = \left( \bA{cccccc} 1 & 1 & 5 & 4 & -11 & 0  \\ 3 & 3 & 4 
& 1 & 0 & -11  \\ \eA \right)\ .
\ee
The structure of this charge matrix is very similar to 
(\ref{Qia2517911}), with five phases, the phase boundaries being
$\phi_1=\phi_2=(1,3), \phi_3=(5,4), \phi_4=(4,1), \phi_5=(-1,0), 
\phi_6=(0,-1)$. The FI-parameters do not run, being marginal. The 
phases, related by marginal deformations, correspond to the unresolved 
orbifold, partial blowup by $R_1$ or $R_3$, and complete blowups by 
$R_1,R_3$, or $R_3,R_1$, the last two related by a 4-dim flop.
After the $R_1$ blowup, the residual singularity 
$\BZ_4 (1,1,-3,1)$ contains the GSO-preserved modulus $P_3\equiv 
({1\over 4},{1\over 4},{1\over 4},{1\over 4})=1$ which completely 
resolves it, while the residual $\BZ_5 (1,1,4,-11)\equiv \BZ_5 
(1,1,1,1)$ singularity is terminal.

The $Q=(\bA{ccccc} 1 & 5 & 6 & -4 & -8 \eA)$ singularity is defined 
by the toric cone with\ $e_1=(-5,-6,4,8),\\ e_2=(1,0,0,0), 
e_3=(0,1,0,0), e_4=(0,0,1,0), e_5=(0,0,0,1)$, in a 4-dim lattice.
The defining relation\ $-5e_2=e_1+6e_3-4e_4-8e_5$ shows the subcone
$C0;e_1,e_3,e_4,e_5)$ to be the orbifold $\BZ_5 (1,1,-4,-8)$. 
The relation\ $(-1,-1,1,2)\equiv e_6={1\over 5} (e_1+e_3+e_4+2e_5)$, 
shows $e_6$ to be an interior lattice point that lies on the 
marginality hyperplane of this orbifold: using (\ref{Rjlattpt}), 
it can be recognized as the GSO-preserved $j=1$ twisted sector 
modulus of $\BZ_5 (1,1,-4,-8)$.
The phase structure of this system, obtained from a GLSM with charge 
matrix
\be\label{Qia15648}
Q_i^a = \left( \bA{cccccc} 1 & 5 & 6 & -4 & -8 & 0  \\ 0 & 1 & 1 
& -1 & -2 & 1  \\ \eA \right)\ ,
\ee
is similar to that of Figure~\ref{figflip}, except that there are 
five phases, the phase boundaries (two of them coincident) being\ 
$\phi_1=(1,0), \phi_2=(5,1), \phi_3=(6,1), \phi_4=\phi_5=(-4,-1), 
\phi_6=(0,1)$. The FI parameters are marginal and have no RG flow.
Being supersymmetric, there are no flips, just marginal flops:
crossing the various phase boundaries corresponds to either a flop 
or condensation of a twisted sector modulus in some residual orbifold 
subcone.

\end{document}